\documentclass[aps,prd,reprint,groupedaddress,showkeys,floatfix]{revtex4-1}

\usepackage{graphicx}
\usepackage{hyperref}
\usepackage{amsmath}
\usepackage{MnSymbol}
\usepackage{cleveref}
\usepackage{subfig}
\usepackage{lineno}
\usepackage{subfiles}
\usepackage{enumitem}

\usepackage{xspace}
\newcommand{\V}{\ensuremath{\vec{v}}\xspace}
\newcommand{\vb}{\ensuremath{\vec{v} \times \vec{B}}\xspace}
\newcommand{\vvb}{\ensuremath{\vec{v} \times (\vec{v} \times \vec{B})}\xspace}

\newcommand{\Xmax}{\ensuremath{X_{\rm max}}\xspace}
\newcommand{\dX}{\ensuremath{d_{\Xmax}}\xspace}
\newcommand{\sina}{\ensuremath{\sin\alpha}\xspace}
\newcommand{\sect}{\ensuremath{\sec\,\theta}\xspace}
\newcommand{\gcm}{\ensuremath{{\rm g}\,{\rm cm}^{-2}}\xspace}

\newcommand{\arel}{\ensuremath{a_{\rm rel}}\xspace}
\newcommand{\aBar}{\ensuremath{\bar{a}_{\rm rel}}\xspace}
\newcommand{\aBarInv}{\ensuremath{\frac{1}{\aBar}}\xspace}

\begin{document}

\title{Simulation Study of the Relative Askaryan Fraction at the South Pole}

\author{Ek Narayan Paudel}
\email{narayan@udel.edu}
\author{Alan Coleman}%
\email{alanc@udel.edu}
\author{Frank G Schroeder}%
 \email{fgs@udel.edu}
 \altaffiliation[also at]{ Institute for Astroparticle Physics, Karlsruhe Institute of Technology, Germany.}
\affiliation{%
Bartol Research Institute, Department of Physics and Astronomy, University of Delaware, Newark DE, USA.
}

\date{\today}

\begin{abstract}
We use \textsc{CoREAS} simulations to study the ratio of geomagnetic and Askaryan radio emission from cosmic-ray air showers at the location of the South Pole. 
The fraction of Askaryan emission relative to the total emission is determined by the polarization of the radio signal at the time of its peak amplitude.
We find that the relative Askaryan fraction has a radial dependence increasing with the distance from the shower axis -- with a plateau around the Cherenkov ring.
We further find that the Askaryan fraction depends on shower parameters like zenith angle and the distance to the shower maximum.
While these dependencies are in agreement with earlier studies, they have not yet been utilized to determine the depth of the shower maximum, $X_\mathrm{max}$, based on the Askaryan fraction. 
Fitting these dependencies with a polynomial model, we arrive at an alternative method to reconstruct $X_\mathrm{max}$ using a measurement of the Askaryan fraction and shower geometry as input. Depending on the measurement uncertainties of the Askaryan fraction, this method provides a measurement of \Xmax with a resolution similar to other methods of reconstructing \Xmax from radio observables. Although existing template methods give superior resolution, they are more computationally intensive. Consequently, the use of polarization to extract the Askaryan fraction of the radio signal should be considered as an additional input observable in future generations of template-fitting reconstruction and other multivariate approaches for measuring \Xmax. 
\end{abstract}
\maketitle



\section{\label{sec:intro}Introduction}
Ultra-high-energy cosmic rays cannot be detected directly by detectors aboard high-altitude balloons and satellites due to their flux being much lower than 1\,m$^{-2}$\,yr$^{-1}$. 
Instead, the air showers that they generate, after interacting with the Earth's atmosphere, are large enough to be observed by ground-based detectors. 
One type of detector for air showers with energies $\gtrsim$\,$10^{16}\,$eV is an antenna arrays, detecting the radio emission of air showers~\cite{Huege:2016veh,Schroder:2016hrv}. 
The radio emission is generally coherent and forward beamed. The main emission mechanisms in air are: 
\vspace{-\topsep}
\begin{description}[leftmargin=0cm]
   \item[Geomagnetic emission] The electrons and positrons present in the electromagnetic component of the air shower are deflected due to the Lorentz force of the Earth's magnetic field inducing a transverse current perpendicular to the shower axis. The time variation of the current leads to electromagnetic radiation which is linearly polarized in the \vb direction, except for very inclined showers with a shower maximum at high altitude where the geomagnetic emission pattern is more complex~\cite{James:2022mea} (the transition starts at $\gtrsim$\,8\,km above sea level for our case, which can be the case for showers at zenith angles $\theta > 60^{\circ}$, i.e., at the edge of the parameter space investigated here). \vspace{-\topsep}
   \item[Askaryan emission] Compton scattering adds electrons to the shower front during shower development. Meanwhile, heavier positive ions lag behind and positrons annihilate. Therefore, a time-varying negative charge excess builds up in the shower front which is responsible for radially polarized radio emission called Askaryan or \emph{charge excess} emission. This emission is typically weaker than Geomagnetic emission, except for the showers nearly parallel to the Earth's magnetic field.
\end{description}

 There are still open questions about the sources of these ultra-high-energy cosmic rays and the processes responsible for accelerating them to ultra-high energies~\cite{Sarazin:2019fjz,Schroder:2019rxp}.
 Understanding the evolution of the mass composition of cosmic rays as a function of energy is crucial for a full understanding of the acceleration mechanisms and possible sources.
 The value of \Xmax, the slant depth into the atmosphere where the number of electromagnetic particles of the air shower is the largest (shower maximum), is dependent on mass as it is related to the cross section of the primary particle's interactions with nucleons in air particles and to the energy per nucleon of the primary particle.
 As the emission in the radio band near \Xmax is maximal~\cite{Glaser:2016qso}, the properties of the radio signal as seen on the ground have dependence on  \Xmax.
 Various parameters of the radio signal have already been shown to contain information on \Xmax, such as the shape of the lateral distribution of the radio amplitude~\cite{LOPES:2012xou,Buitink:2014eqa,Bezyazeekov:2018yjw}, the wave front~\cite{Apel:2014usa,Corstanje:2014waa}, or the slope of the frequency spectrum~\cite{Grebe:2013lvs}. 
 As we show in this paper, the fraction of the Askaryan emission relative to the geomagnetic one, which is accessible from the polarization of the radio signal, is another parameter sensitive to \Xmax. 
Complementing earlier studies on the Askaryan fraction had a different focus~\cite{PierreAuger:2014ldh,Corstanje:2015icrc,Kostunin:2015taa}. Among other dependencies they had shown a dependence of the radio emissions on the air density at the shower maximum~\cite{Glaser:2016qso}, but have not further investigated the dependence on \Xmax, , i.e., the atmospheric depth of the shower maximum.
Our simulation study explicitly shows the dependence on the highly relevant observable \Xmax and investigates whether the Askaryan fraction can serve as a tool to potentially measure \Xmax with polarization-sensitive radio arrays.
 
 In this paper, we perform a \textsc{CoREAS} simulation study of air showers at the South Pole to describe the relative emission at higher frequencies than previously studied ($\leq$\,80\,MHz).
 This is motivated by the upcoming enhancement of the existing IceTop array, the surface component of the IceCube Neutrino Observatory~\cite{IceCube:2016zyt,IceCube:2012nn}. The enhancement will include adding scintillators and, more relevant to this work, antennas which will be used in the 70--350\,MHz band~\cite{roxmarieProceeding}. The addition of radio antennas at IceTop will allow for a pure measurement of the electromagnetic content of air showers and in particular the mass-proxy \Xmax. However, the technique described in this paper can be applied to any radio array that uses the same frequency band. For any such array, interpreting the expected measurements accurately requires a detailed understanding of the observed radio signal, which motivated this simulation study on the various dependencies of the relative strengths of the Askaryan to the geomagnetic emission.
 
 The dependence of the relative Askaryan fraction of the radio emission with various air shower parameters like the zenith angle, \Xmax, and the distance to the shower maximum (\dX) is described. 
  Section \ref{sec:sim} describes the simulation of the radio emission from the cosmic-ray air showers that were used in this analysis. 
 Section \ref{sec:relAmp} describes the dependence of the geomagnetic emission on the geomagnetic angle of the shower, $\alpha$ and the method we used for separating individual contribution of the two radio emission mechanism and calculating the relative Askaryan fraction. It further describes the parameterization of the relative Askaryan fraction with \dX.
 Finally, we discuss the relevance and potential impact of our findings on \Xmax measurements with future radio arrays.
\section{\label{sec:sim}Simulation of the radio emission}
\textsc{CORSIKA} with the \textsc{CoREAS} extension was used to simulate cosmic ray air showers and the radio emission from the charged particles in the shower~\cite{Heck:1998vt,Huege:2013vt}. 
We simulated 1800 proton and iron initiated air showers each, with the primary energy assigned according to, $d N/d E \propto E^{-1}$, from $10^{17.0}$\,eV to $10^{17.1}$\,eV, with zenith angles distributed according to an isotropic flux, $dN/d\theta \propto \cos \theta\,\sin\theta$, with $\sin ^{2}\theta$ from 0.0--0.9, and with a random azimuth angle. 
We used an observation level of 2840\,m above sea level, magnetic field of strength $|\vec{B}|$ = 54.58\,$\mu$T inclined at an angle of $17.87^{\circ}$ from the zenith and the average April South Pole atmospheric profile (atmosphere model 33 in \textsc{CORSIKA}). 
\textsc{FLUKA2011}~\cite{Bohlen:2014buj} and \textsc{SIBYLL} 2.3d~\cite{PhysRevD.102.063002} were used as the high- and the low-energy hadronic interaction models and the \textsc{CORSIKA} thinning algorithm (set to $10^{-6}$) was used to speed up the computationally expensive simulations.

\begin{figure}[t]
    \centering
    \includegraphics[width=8.6cm]{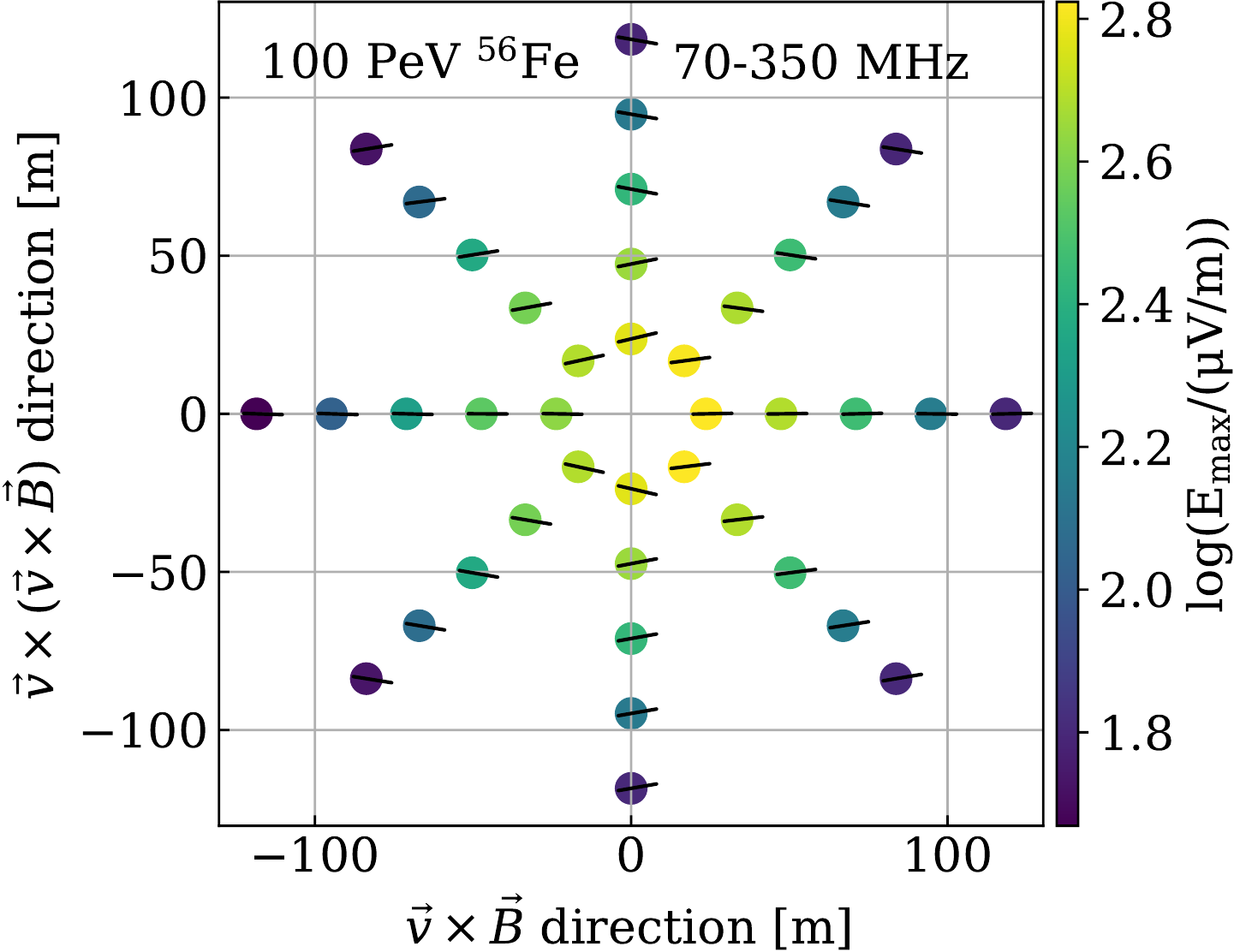}
    \caption{Map of the electric field amplitudes, given by the color, for a $100\,$PeV vertical iron shower at ground at the South Pole. The lines represents the polarization planes of the electric fields at the signal peak. For reference, the IceTop Surface Enhancement will span over a radius of $\sim$\,600\,m.} 
    \label{fig:antennaMap}
\end{figure}

We used a star-shaped layout of sampling locations, similar to one shown in \cref{fig:antennaMap}, as an input to \textsc{CoREAS}, and simulated the radio emission from the air shower at each of those locations. 
The star shape we used has 8 spokes with 20 sampling locations along each spoke totaling 160 such sampling points. 
The distance between those sampling locations along the spokes is adaptive and is based on the location of Cherenkov ring center. 
For more inclined showers, the spacing between the sampling points is higher and the star pattern extends to a larger area to be able to cover the larger radio footprint without increasing the total number of sampling points and hence limiting the required computational time. 

\begin{figure}[t]
    \centering
    \includegraphics[width=8.6cm]{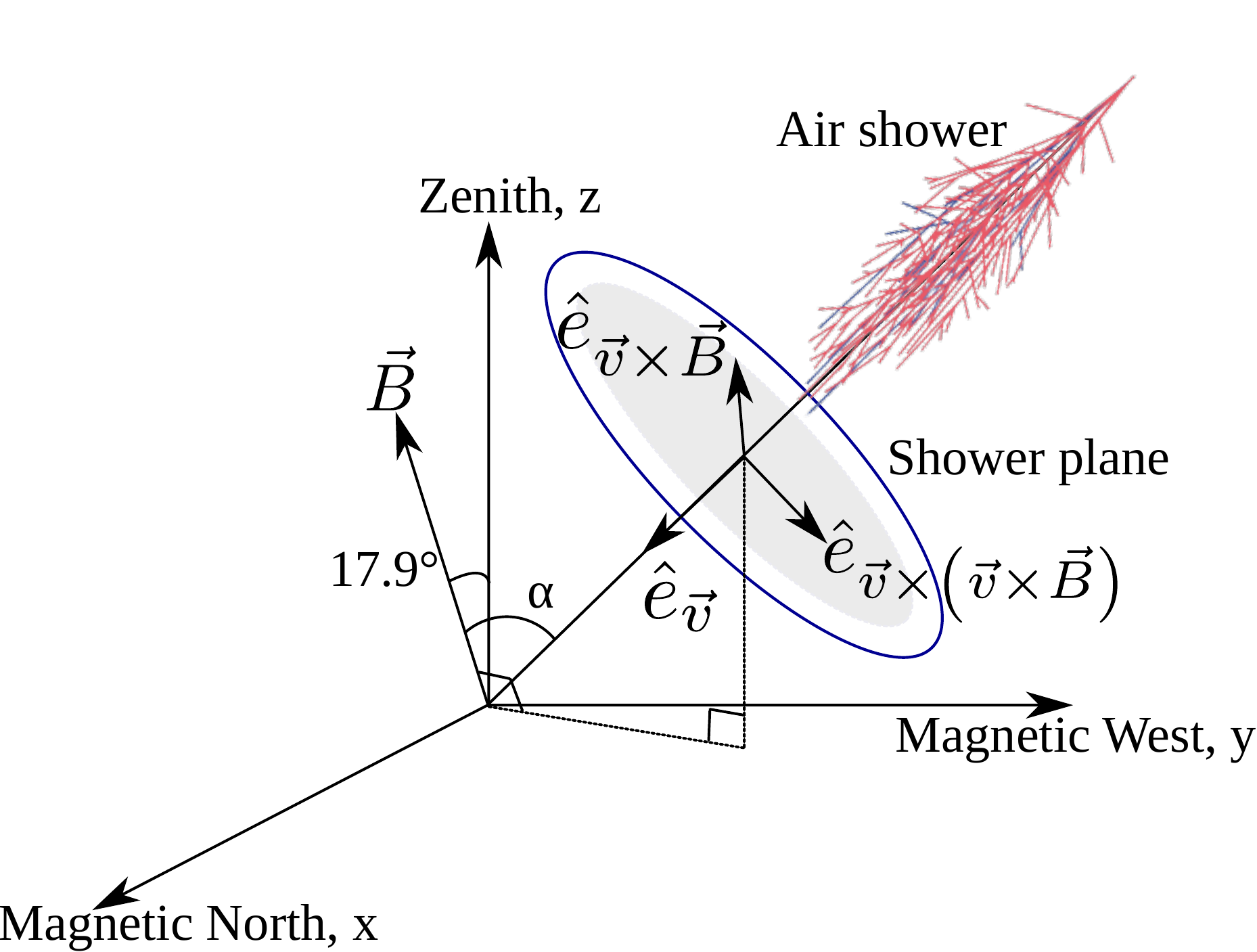}
    \caption{The shower coordinate system used in this study with respect to the magnetic coordinate system (image from~\cite{Paudel:2021dpn}).}
    \label{fig:showerCoordinate}
\end{figure}

When studying radio emission, we use the coordinate system as defined in \cref{fig:showerCoordinate} with one axis aligned along the \vb direction in the shower plane and the other along the \vvb direction.
In this coordinate system, the geomagnetic emission is aligned along the \vb axis, making it easier to separate from the Askaryan emission using the polarization information.
For this reason, we also ensured that the star pattern included spokes that were along the \vb and \vvb directions, as shown in \cref{fig:antennaMap}.

\begin{figure}[t]
    \subfloat[\centering]{\includegraphics[width=8.6cm]{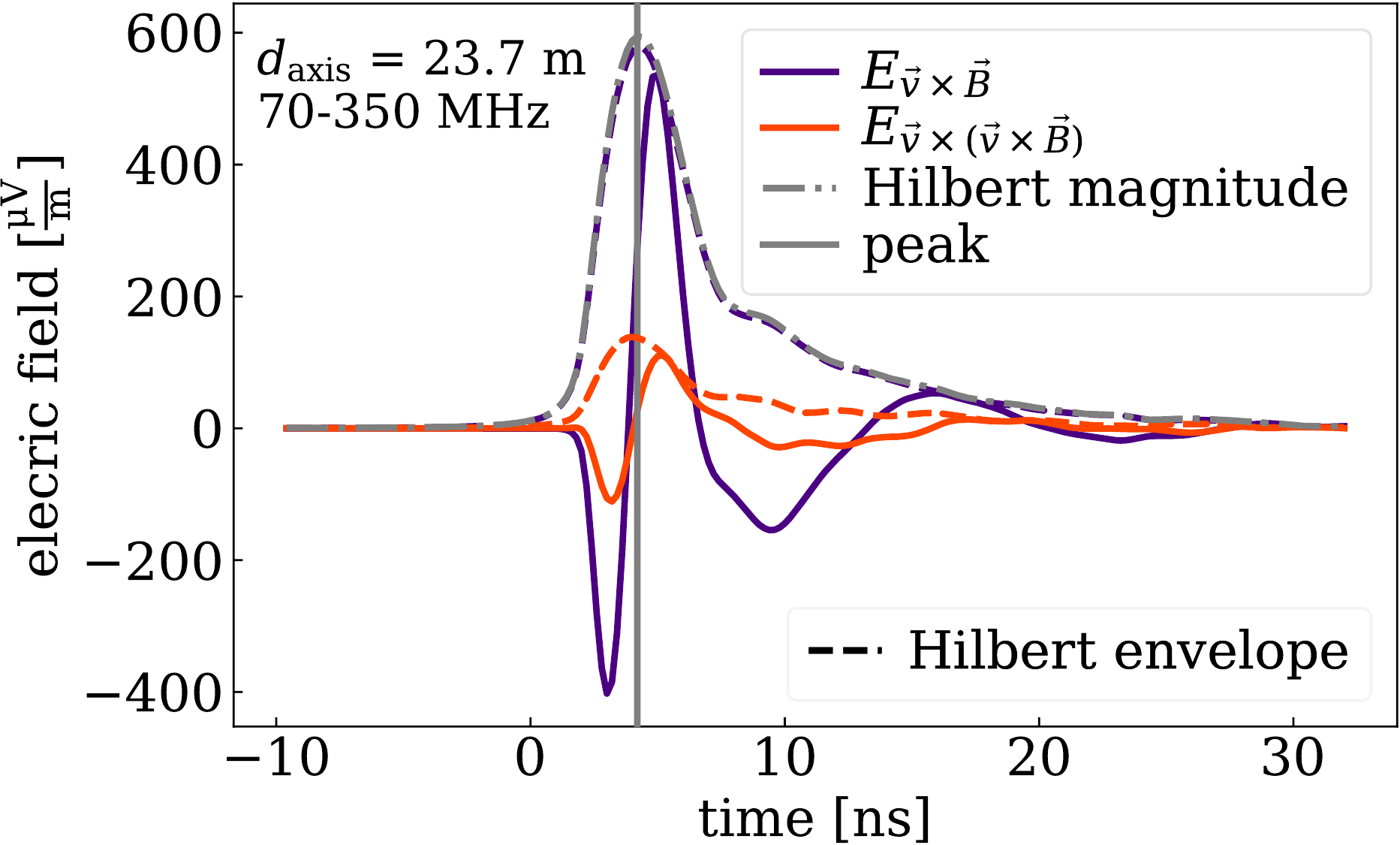}\label{fig:trace42}}
    \qquad
    \subfloat[\centering]{\includegraphics[width=8.6cm]{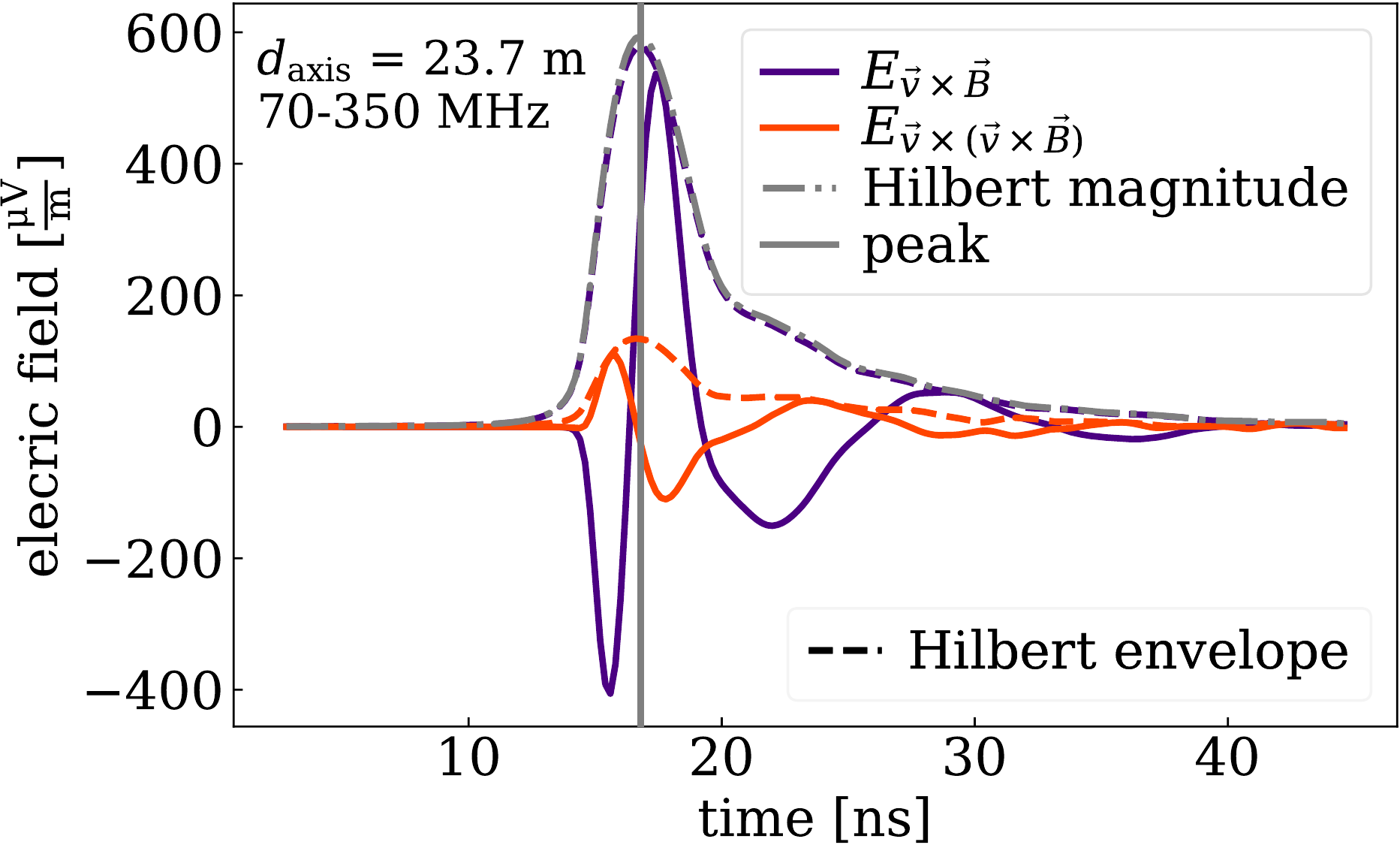}\label{fig:trace122}}
    \caption{Simulated time traces and the respective Hilbert envelopes for a location along the \protect\subref{fig:trace42} positive \vvb axis and \protect\subref{fig:trace122} negative \vvb axis. Hilbert magnitude represents the global Hilbert envelope.
    }
    \label{fig:trace42122}
\end{figure}

The polarization of the radio emission for each location was determined from the waveforms.
\Cref{fig:trace42122} shows example waveforms at two sampling locations on the positive and negative \vvb axis, respectively, after filtering them in the frequency band of 70 to 350\,MHz. 
The Hilbert envelope was used to identify the time of peak emission, indicated by the vertical grey line.
The amplitudes of the electric field vectors along the \vb and \vvb axes at that time are taken to define the plane of polarization.
For each sampling location in \cref{fig:antennaMap}, the polarization plane is shown by the black line.

\begin{figure}[t]
    \centering
    \includegraphics[width=7cm]{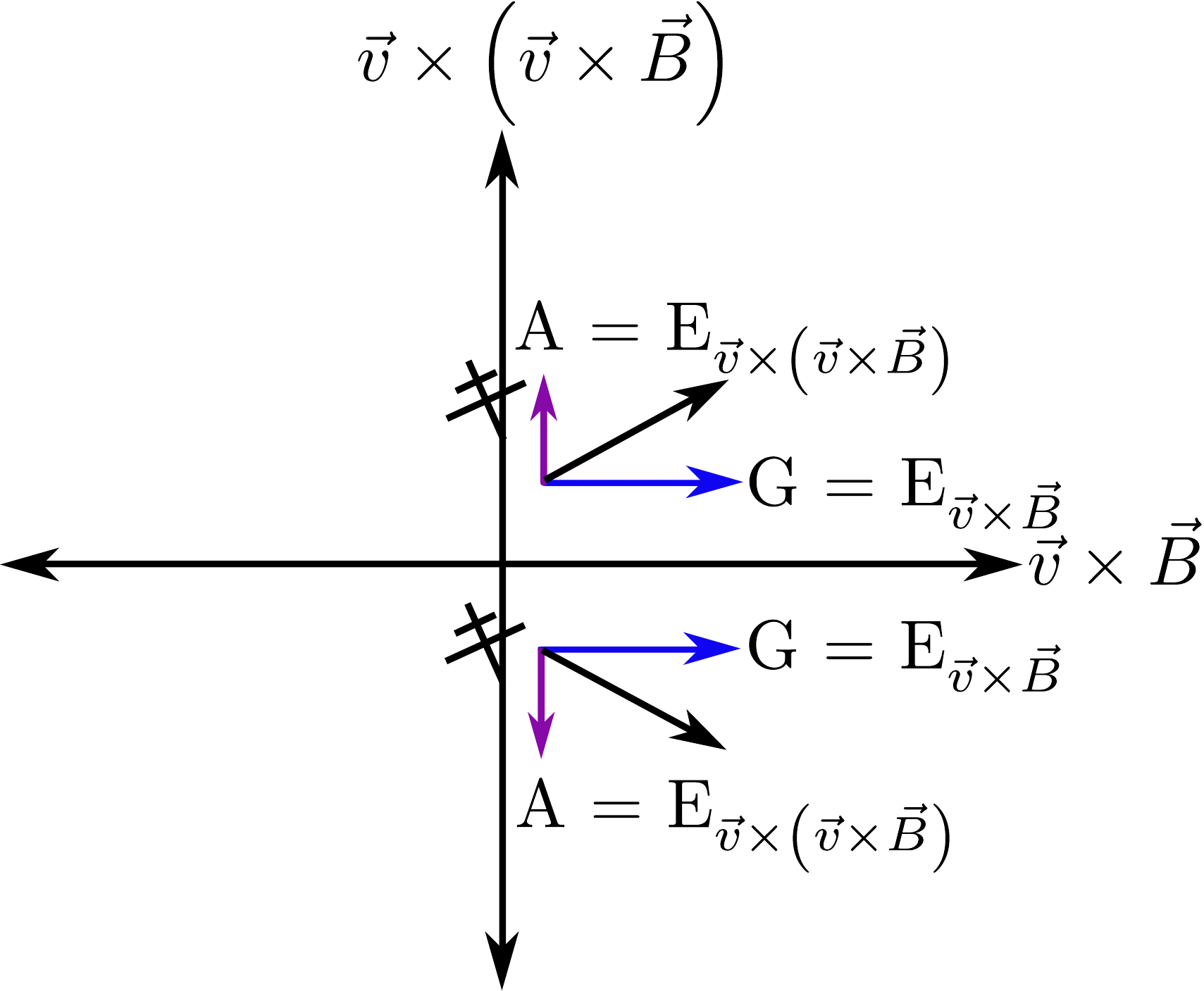}
    \caption{The figure shows how the polarization of the radio emission at the Hilbert peak of the signal appears at the radio sampling locations along the \vvb axis. }
    \label{fig:caseAG}
\end{figure}

For sampling locations along the \vvb axis, the geomagnetic and Askaryan components (hereon referred to as $G$ and $A$, respectively) are orthogonal to each other as depicted in \cref{fig:caseAG}.
So, for these locations, the component of the electric field along the \vb axis, $\mathbf{E}_{\vb}$, represents the geomagnetic emission and the component along the \vvb axis, $\mathbf{E}_{\vvb}$, represents the Askaryan emission
(because the Poynting vector defining the wavefront shape deviates by only about $1^{\circ}$ from the shower axis {\V}~\cite{Apel:2014usa}, the $\mathbf{E}_{\V}$ component of the electric field is of the order of $1\,\%$ of the total field and is neglected in this study).
Thus, polarization measurements along the \vvb axis provide one clear method to determine the relative fraction of the Askaryan emission.

We note that the method of calculating the polarization at the maximum of the global Hilbert envelope of the electric field assumes that the geomagnetic and Askaryan emission peak together in phase. 
This is another approximation which neglects the circular polarization component of the radio emission of air showers~\cite{Scholten:2016gmj}.
The validity of this approximation at the location of the Cherenkov ring region, which is  used in our study, is further described in the appendix \ref{appen:phaseShift}. 

For locations in the shower plane along the \vb axis, the geomagnetic and Askaryan emission components are aligned either in the same direction, superimposing constructively, or in the opposite direction, superimposing destructively, on the sampling locations along the two spokes of the \vb axis creating an asymmetry in amplitude of the radio emission as is apparent from the color in \cref{fig:antennaMap}.
Consequently, the asymmetry of the lateral distribution of the radio amplitude along the \vb axis provides another method to determine the Askaryan emission.
%
%

Both methods generally yield consistent results on the Askaryan fraction (see appendix \ref{appen:alternateMethod}), but the asymmetry methods comes with the complication that the amplitude to one side can be zero if the Askaryan and geomagnetic emission are of equal strength. 
Therefore, we have decided to use the polarization method to determine the Askaryan fraction in this study.

At observation angles larger than the Cherenkov angle, the radio emission can become incoherent, and then usually is below the detection threshold of contemporary experiments.
Hence, for this study, we aim to characterize the region of the radio footprint on the ground only where there is coherent emission.
We used a signal-to-noise ratio (SNR) cut of ${\rm SNR}>10^{4}$ in the pure \textsc{CoREAS} simulations to exclude the locations where the waveforms with weak, non-coherent signals are present where,
\begin{equation}
    \label{eq:SNR}
    {\rm SNR} =\left(\frac{{\rm S}_{\rm peak}}{{\rm N}_{\rm RMS}}\right)^2.
\end{equation}
Here, ${\rm S}_{\rm peak}$ is the peak of the Hilbert envelope of the signal trace and ${\rm N}_{\rm RMS}$ is the root mean square of the last $40\,\%$ of the waveform.
This SNR cut was applied uniformly across all simulated waveforms in the library, after applying the bandpass filter.
\section{\label{sec:relAmp}Relative amplitude of the emission processes}

\begin{figure}[t]
    \centering
    \includegraphics[width=8.6cm]{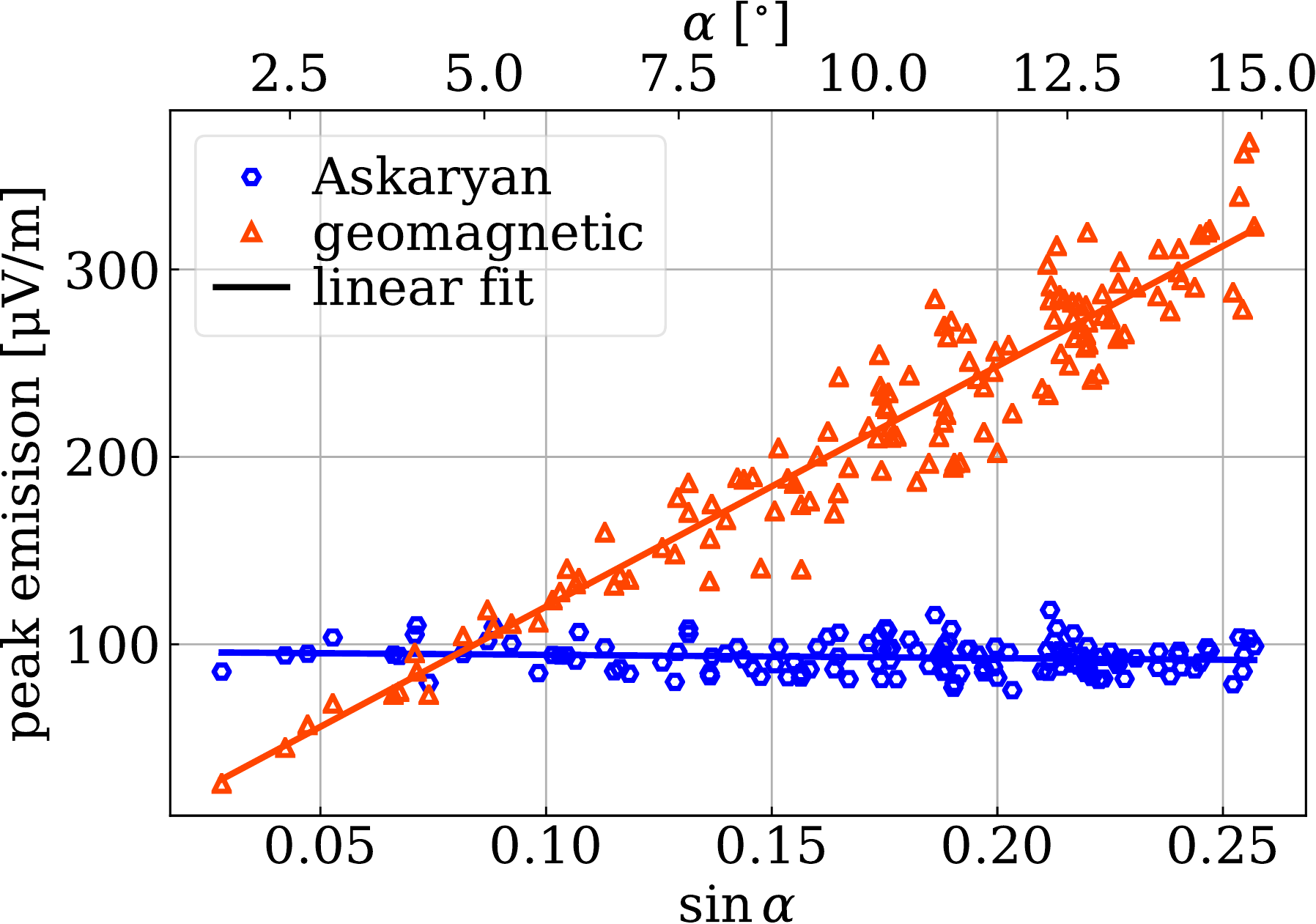}
    \caption{Amplitudes of the Askaryan and the geomagnetic emission at $\simeq$\,23\,m from the shower axis in the shower plane for various \sina simulated for the iron showers of $\sim$\,100\,PeV primary energy.} 
    \label{fig:sinInt}
\end{figure}

Of the two processes, the Askaryan emission is expected to be roughly constant in amplitude for all arrival directions.
However, the dependence of the geomagnetic emission on the Lorentz force results in a strength that scales with the sine of the angle with the geomagnetic field, \sina.
In \cref{fig:sinInt}, the peak geomagnetic and Askaryan amplitudes at the Hilbert maximum  of the simulated radio signal at $\simeq$\,23\,m from the shower axis for various iron showers are plotted against \sina. 
We find that the geomagnetic emission is weaker than the Askaryan emission for showers with $\alpha$ $\lesssim$ 5$^{\circ}$. 
Also, the expected linear dependence of G on \sina is seen.
Taking into account this first-order dependence of the geomagnetic amplitude, we define the relative Askaryan fraction as,
\begin{equation}
    \label{eq:AFvvB}
    \arel \equiv \sin\,\alpha \left ( \frac{A }{G} \right ).
\end{equation}
Thus, $a_{\rm rel}$ corresponds to the relative strengths of the Askaryan emission for showers with maximum geomagnetic emission. 
In the rest of this work, we study the dependence of the quantity $a_{\rm rel}$ on various air shower parameters.

\subsection{The radial dependence of the relative Askaryan fraction}

The value of \arel as a function of axial distance along the \vvb axis was  studied first.
For each of the locations which passed the SNR cut, we estimated the uncertainty in \arel using the bootstrap method. 
In each waveform, there is some level of `noise' which may be due to physical properties such as incoherent emission, but may also be artifacts of the simulation, such as from the thinning algorithms used in \textsc{CORSIKA}.
Based on the assumption that this noise is prevalent throughout the entire waveform after the radio pulse, we estimate the uncertainty for each waveform individually.
As with the construction of the SNR values, the RMS of the last 40\% of the waveforms was used as metric for the noise level because the radio pulse simulated by \textsc{CoREAS} is close to the beginning of the trace. 
A three dimensional vector with a random direction and  magnitude equal to the RMS was chosen and then added to the peak electric field before calculating \arel using \cref{eq:AFvvB}. 
After repeating this process 1000 times and combining the values for both the positive and negative \vvb axis, the 68\% interval of the resulting distribution was used as the uncertainty estimate for the mean value of \arel.

\begin{figure}[tb]
    \subfloat[\centering]{\includegraphics[width=8.6cm]{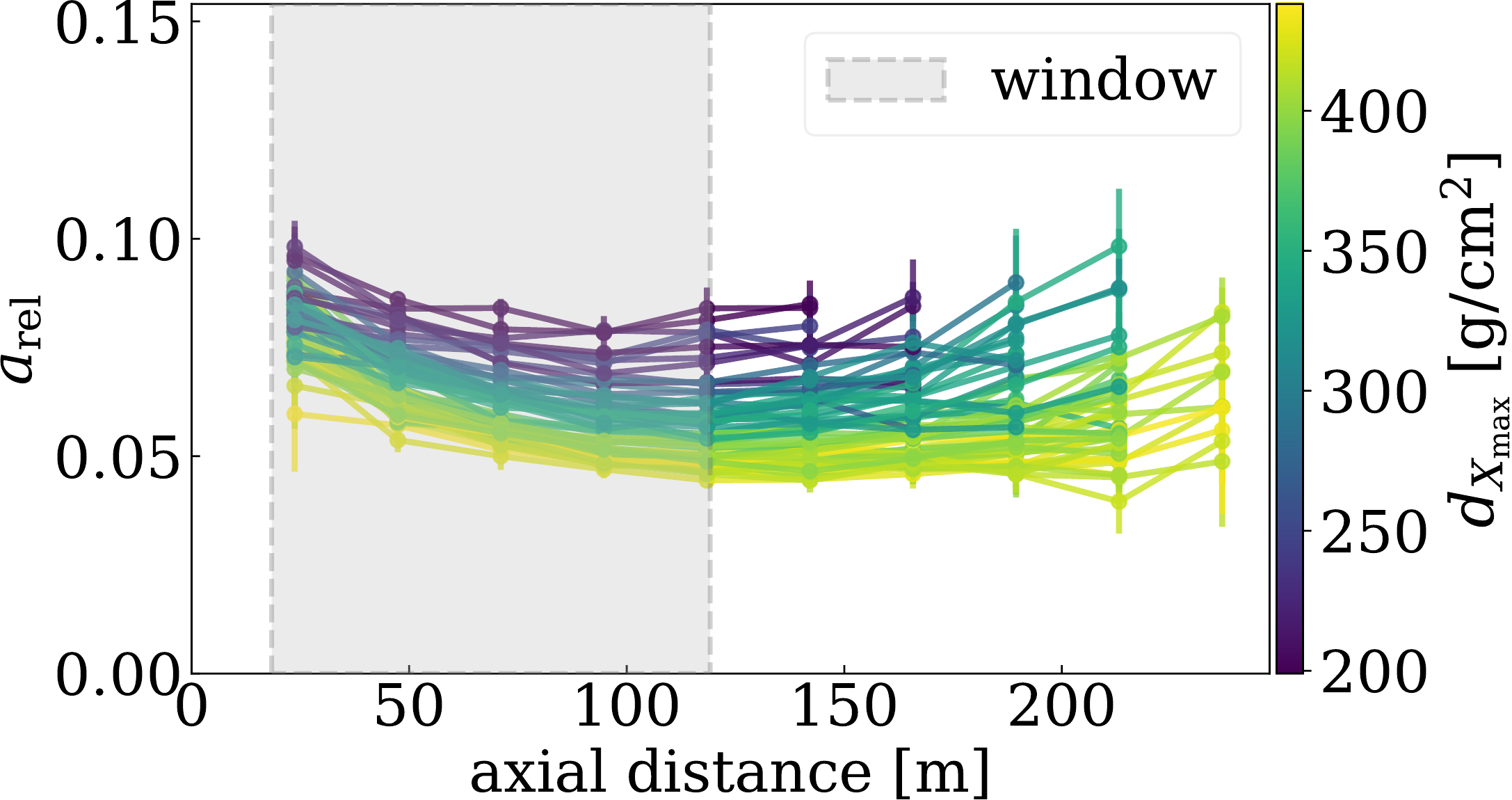}\label{fig:afZen45_46}}
    \qquad
    \subfloat[\centering]{\includegraphics[width=8.6cm]{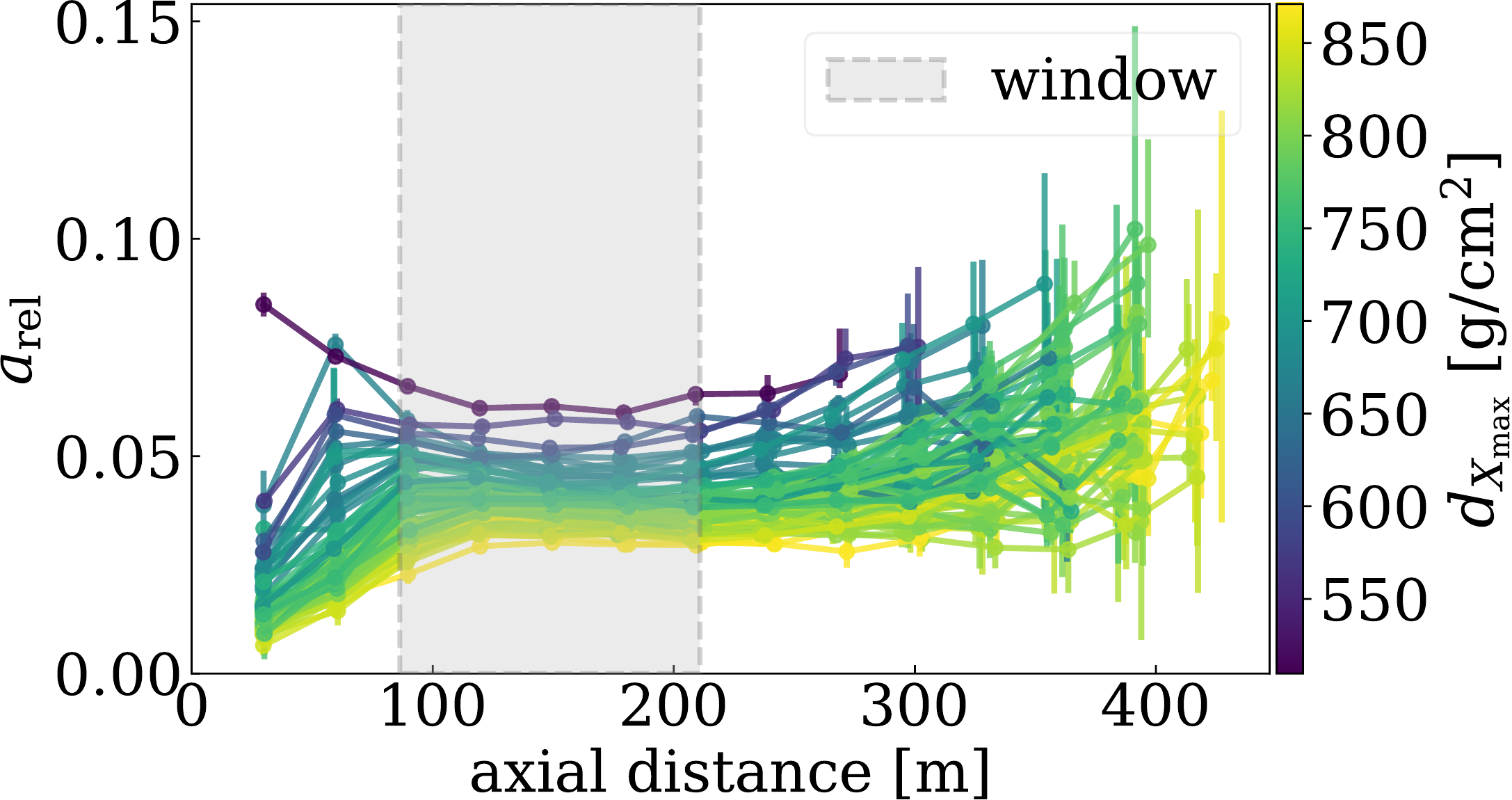}\label{fig:afZen60_61}}
    \caption{ The relative Askaryan fraction plotted against the distance from the shower axis for a mix of proton and iron primaries of energy between 10$^{17.0}$\,eV--10$^{17.1}$\,eV and zenith angles between  \protect\subref{fig:afZen45_46} 45$^{\circ}$--46$^{\circ}$ and \protect\subref{fig:afZen60_61} 60$^{\circ}$--61$^{\circ}$.}
    \label{fig:afZen}
\end{figure}

Examples of the values of \arel for showers at two zenith angle ranges are shown in \cref{fig:afZen}. 
The color of each shower represents the distance to \Xmax, \dX.
Even within the very narrow ranges of zenith angle in each of the two plots, a clear dependence on \dX is observed.
For showers which develop closer to the array, the average values of \arel are larger.
There is another feature which is evident at high zenith angles where, near the shower axis, the Askaryan fraction is particularly small.
This is consistent with radially polarized emission, which must cancel out at the shower axis and has been seen in other simulation studies~\cite{Alvarez-Muniz:2014wna,Glaser:2018byo}.
Further, the orthogonal distance from the shower axis at which the values of \arel level off also has some \dX dependence.
However, this dependence means that a measurement of \arel can be used as a parameter to estimate \Xmax.

\subsection{The value of \texorpdfstring{\arel}\, near the Cherenkov ring}
\label{subsec:ChRng}

The dependence of \arel on \dX requires additional investigation.
However, the non-trivial dependence with axial distance makes quantifying the relative Askaryan fraction impossible with a single value.
Instead, we studied the average value of \arel only near the Cherenkov ring.
This is motivated by the fact that the radio emission is most prominent on the ring and thus this is the most likely place that an experiment would be able to make precise measurements of the polarization.

The location of the Cherenkov ring was studied statistically using the simulated air showers.
For each simulation, the lateral distribution of electric field strength, characterized by the peak of the Hilbert envelope, ${\rm S}_{\rm peak}$, was fit to the sum of an exponential and a Cauchy distribution,
\begin{equation}
    \log ({\rm S}_{\rm peak}(r)) = \frac{a_1}{1 + \left(\frac{r - r_0}{\Delta}\right)^2} + a_2\, e^{-r / \lambda}.
\end{equation}
From each fit, the center and the width of the Cherenkov ring were taken as $r_0$
and $\Delta$. Because this fit was performed on the
logarithm of the field strength, the radial values spanned
by $\pm\Delta$ represents 80--90\,\% of the total energy in the radio band.
\begin{figure}[tb]
    \centering
    \includegraphics[width=8.6cm]{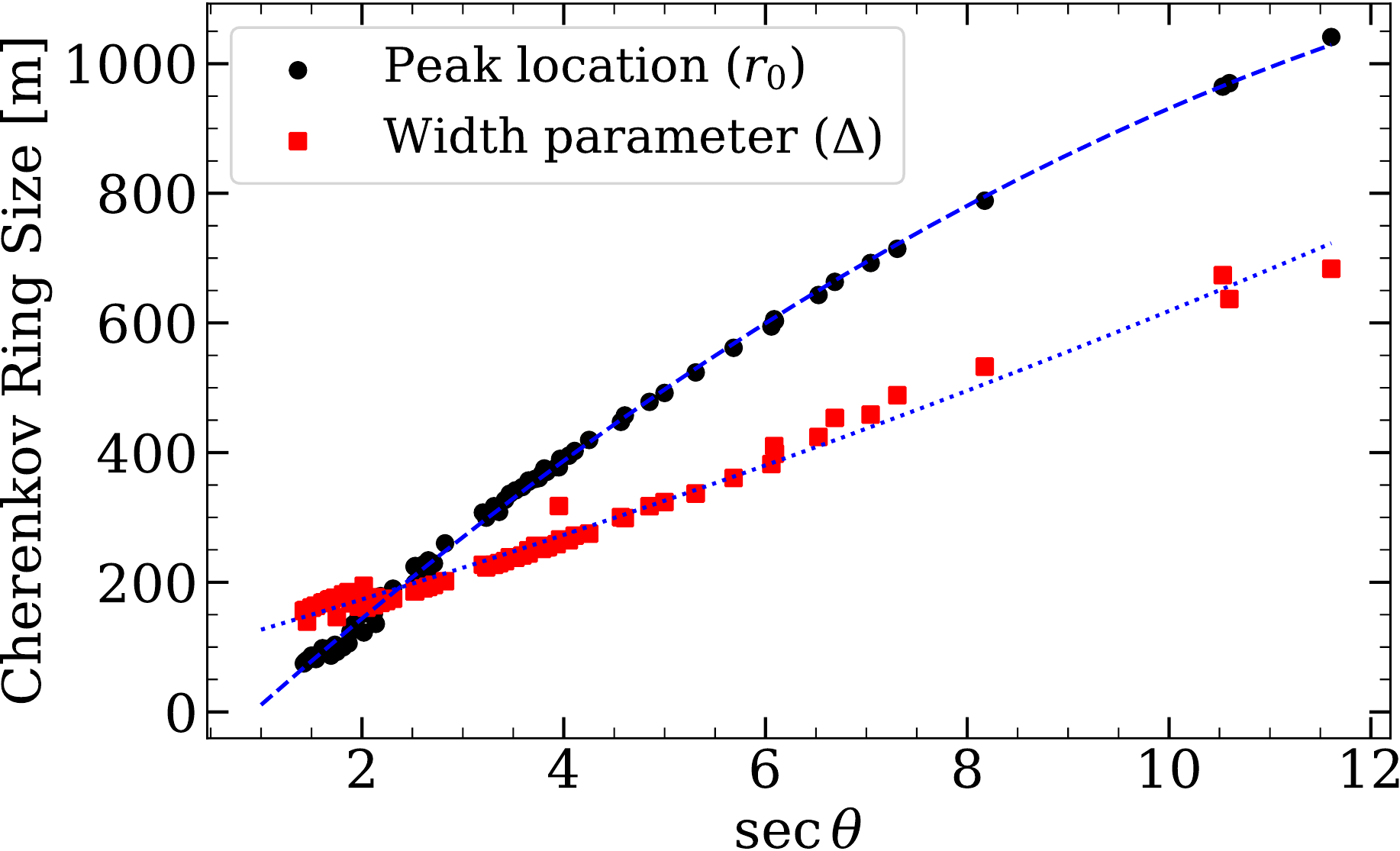}
    \caption{Peak location (black circles) and width parameter (red square) of the Cherenkov ring for showers of various zenith angles. The fit is represented by corresponding lines to the scatter plot.} 
    \label{fig:chrngWindow}
\end{figure}
\Cref{fig:chrngWindow} shows the fit of $r_0$ and $\Delta$ as a function of $\sec \theta$, meant to approximately encapsulate the scaling of the ring size with \dX.
It is clear that for the most vertical showers with $\sec{\theta}\lesssim 2$, the value of $r_0$ is comparable to $\Delta$, in which case the Cherenkov ring overlaps the shower axis.
We fit the values of the center and spread of the ring to second order polynomials,
\begin{equation}
    \label{eq:ChRngFit}
    \begin{split}
        & r_0(\theta) = -3.883 \sec^2 \theta + 145.0 \sec \theta - 130.3\\
        & \Delta(\theta) = 0.9777 \sec^2 \theta + 43.88 \sec \theta + 81.86
    \end{split}
\end{equation}
To avoid characterizing the behavior near the shower axis where $A\longrightarrow 0$, we define a window about the Cherenkov ring to perform the study, given by $r_0 \pm \Delta/3$.
The respective windows are shown in gray for the two zenith angle bands in \cref{fig:afZen}.

To quantify the average value of \arel for a given shower near the Cherenkov ring, we calculate the average value, \aBar, in the window taking into account the uncertainties estimated using the bootstrap method.
In the following sections we will characterize the dependence of \aBar on the various shower parameters.

\subsection{\label{subsec:aRELDep}Dependence of \texorpdfstring{\aBar} \, on air-shower parameters}

\begin{figure}[tb]
    \centering
    \includegraphics[width=\columnwidth]{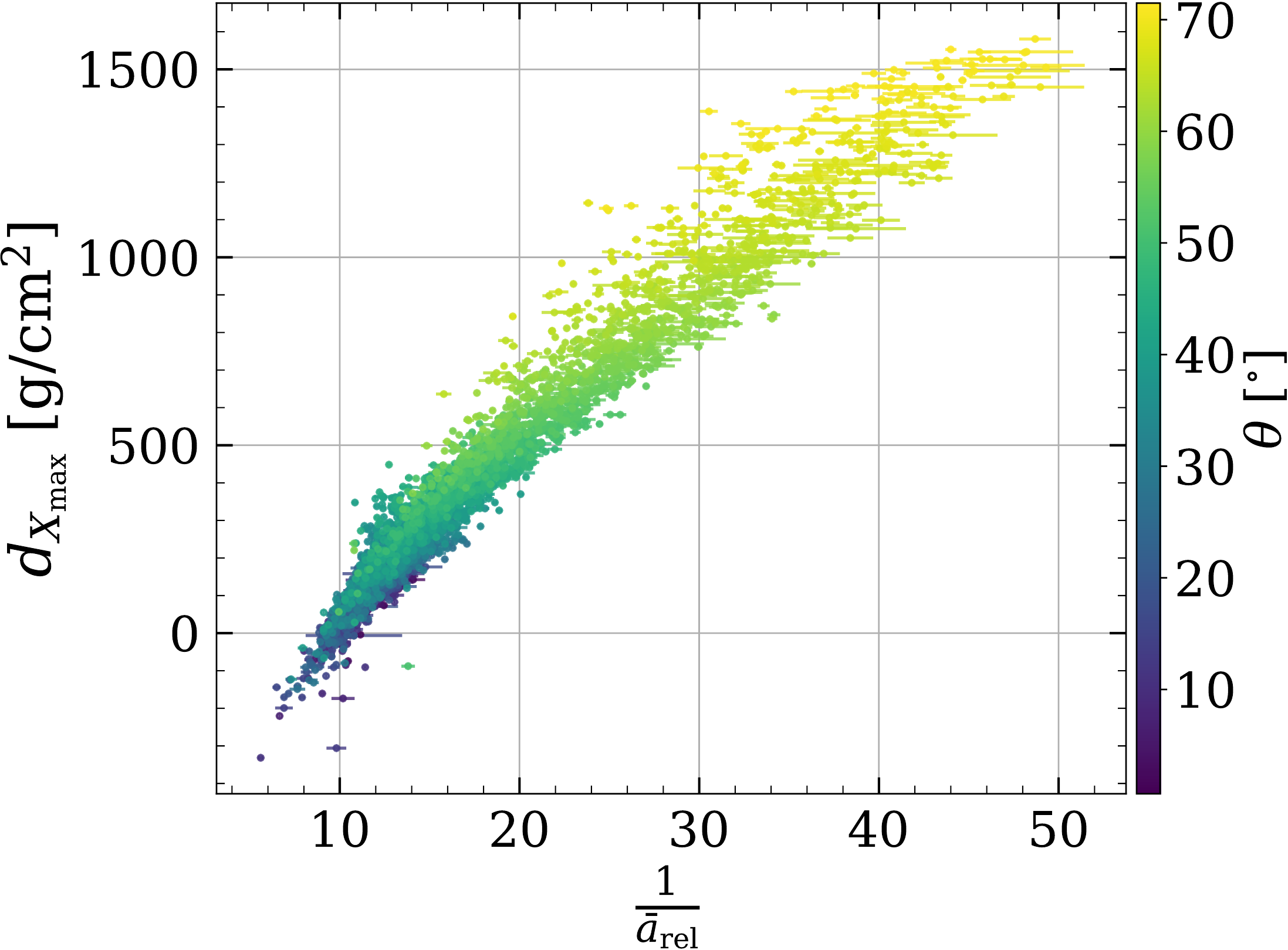}
    \caption{Relationship between the average Askaryan fraction, \aBarInv, and \dX for simulations of a mix of proton and iron primaries with energy between 10$^{17.0}$--10$^{17.1}$\,eV and different zenith angles $\theta$.} 
    \label{fig:afWZenith}
\end{figure}

There was already some evidence, such as in \cref{fig:afZen}, that the value of the Askaryan fraction \aBar cannot be described by a single air-shower parameter, such as \dX.
To understand the full picture, the values \aBar for all of the simulated showers are shown in \cref{fig:afWZenith}.
There is an approximately linear relationship between \dX and \aBarInv.
However, there is also a non-trivial dependence on zenith angle, as indicated by the color of the points.
A relationship between zenith angle $\theta$ and \dX is expected from the increased slant depth to the top of the atmosphere at higher angles.
However, looking at a narrow window in \dX, the variance has very clear zenith angle dependence where the largest values of \aBarInv in that window are for more vertical showers.

We use an empirical model to fit the dependence of these three quantities.
With the ultimate goal of using measurable quantities to determine \Xmax (or equivalently \dX), we chose an empirical function which is a function of the air-shower observables, \aBarInv and \sect,
\begin{equation}
    \small
    \dX = \begin{bmatrix}
    1 &\aBarInv &\left(\aBarInv\right)^2 &\left(\aBarInv\right)^3
    \end{bmatrix} 
    \begin{bmatrix}
    a & b \\
    c & d\\
    e & f\\
    g & h
    \end{bmatrix}
    \begin{bmatrix}
    1 \\ \sect
    \end{bmatrix}.
    \label{eq:dX_pars}
\end{equation}
This equation was fit using the distribution shown in \cref{fig:afWZenith} via a minimized log-likelihood.
The best fit values of $a,\, b,\,c,\,...$ in the matrix above are given in \cref{tab:dX_pars}.
\begin{table}[tb]
    \centering
    \begin{tabular}{c|l}
        $a$ & -1050\phantom{.0000} $\pm$ 10 \\
        $b$ & \phantom{-0}220\phantom{.0000} $\pm$ 10 \\
        $c$ & \phantom{-0}131\phantom{.0000}  $\pm$ \phantom{0}1 \\
        $d$ & \phantom{000}-7.6\phantom{000}  $\pm$ \phantom{0}0.1 \\
        $e$ & \phantom{000}-5.54\phantom{00} $\pm$ \phantom{0}0.03\\
        $f$ & \phantom{-000}1.11\phantom{00} $\pm$ \phantom{0}0.02\\
        $g$ & \phantom{-000}0.072\phantom{0} $\pm$ \phantom{0}0.001 \\
        $h$ & \phantom{000}-0.0187 $\pm$ \phantom{0}0.0003
    \end{tabular}
    \caption{Parameters for the model of \dX given by \cref{eq:dX_pars}. The correlation matrix is given in \cref{app:correlation_matrix}.}
    \label{tab:dX_pars}
\end{table}

\subsection{Attainable resolution of \Xmax}

\begin{figure}[!ht]
    \includegraphics[width=8.6cm]{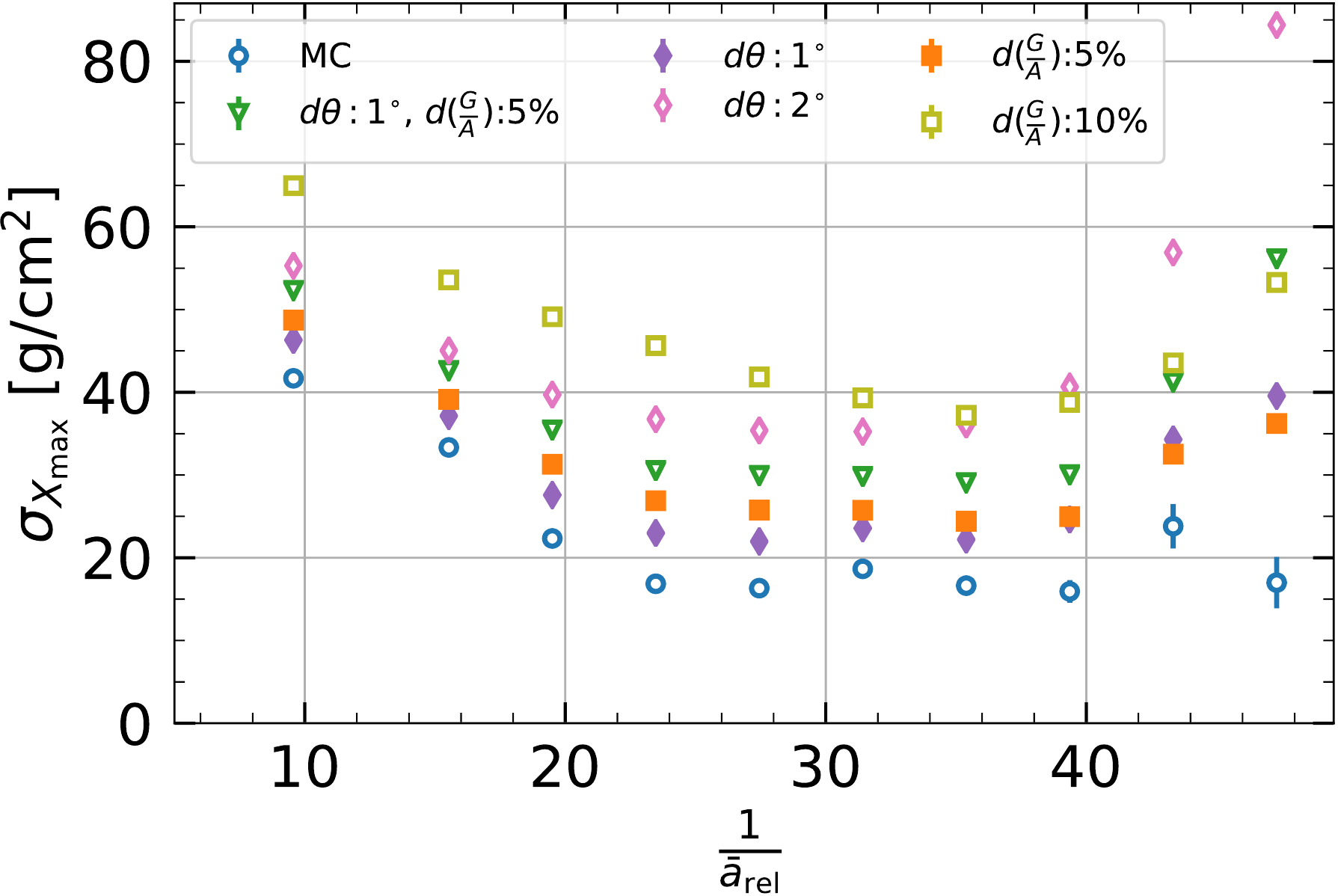} \\
    \includegraphics[width=8.6cm]{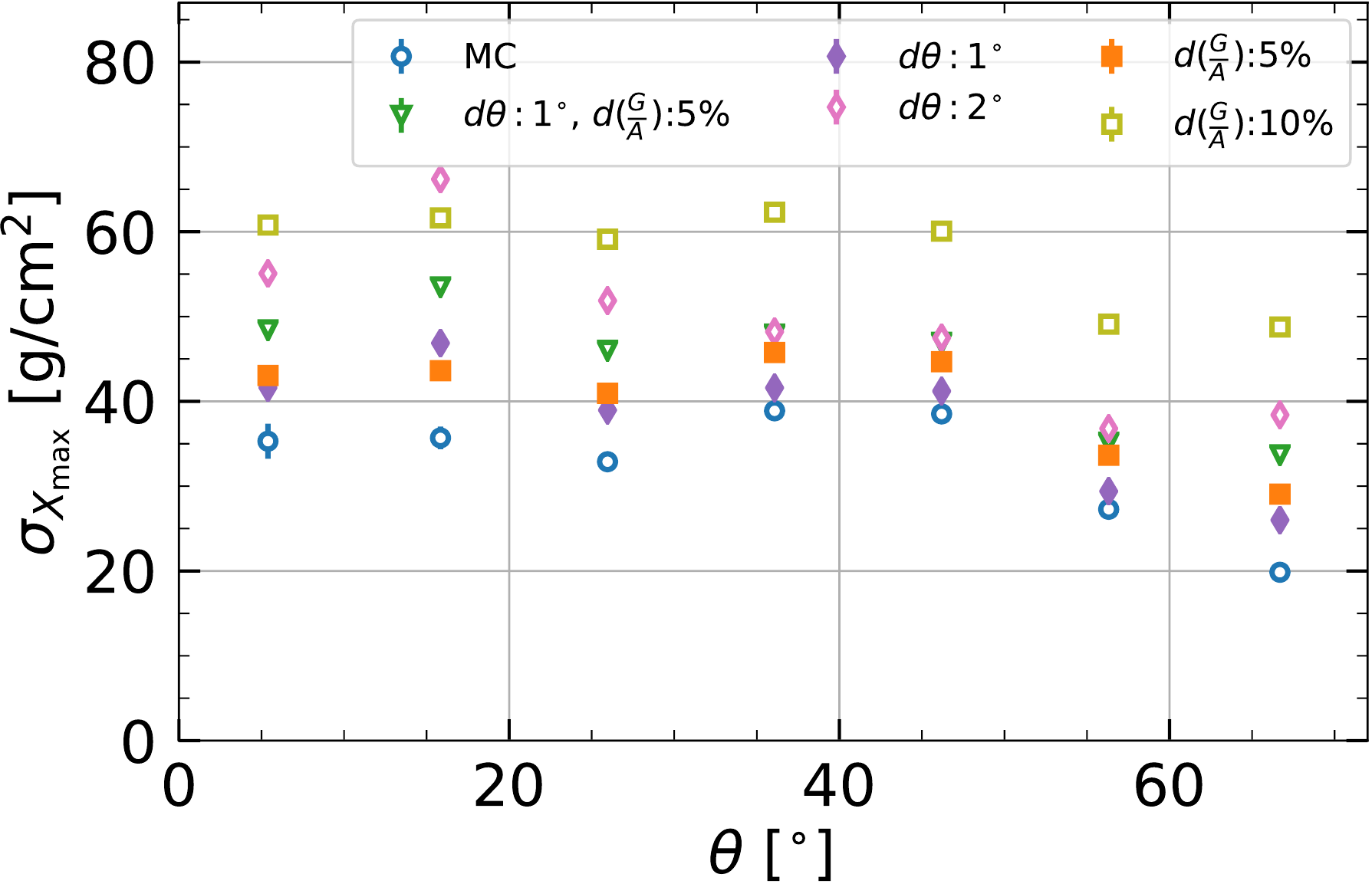}\\
    \includegraphics[width=8.6cm]{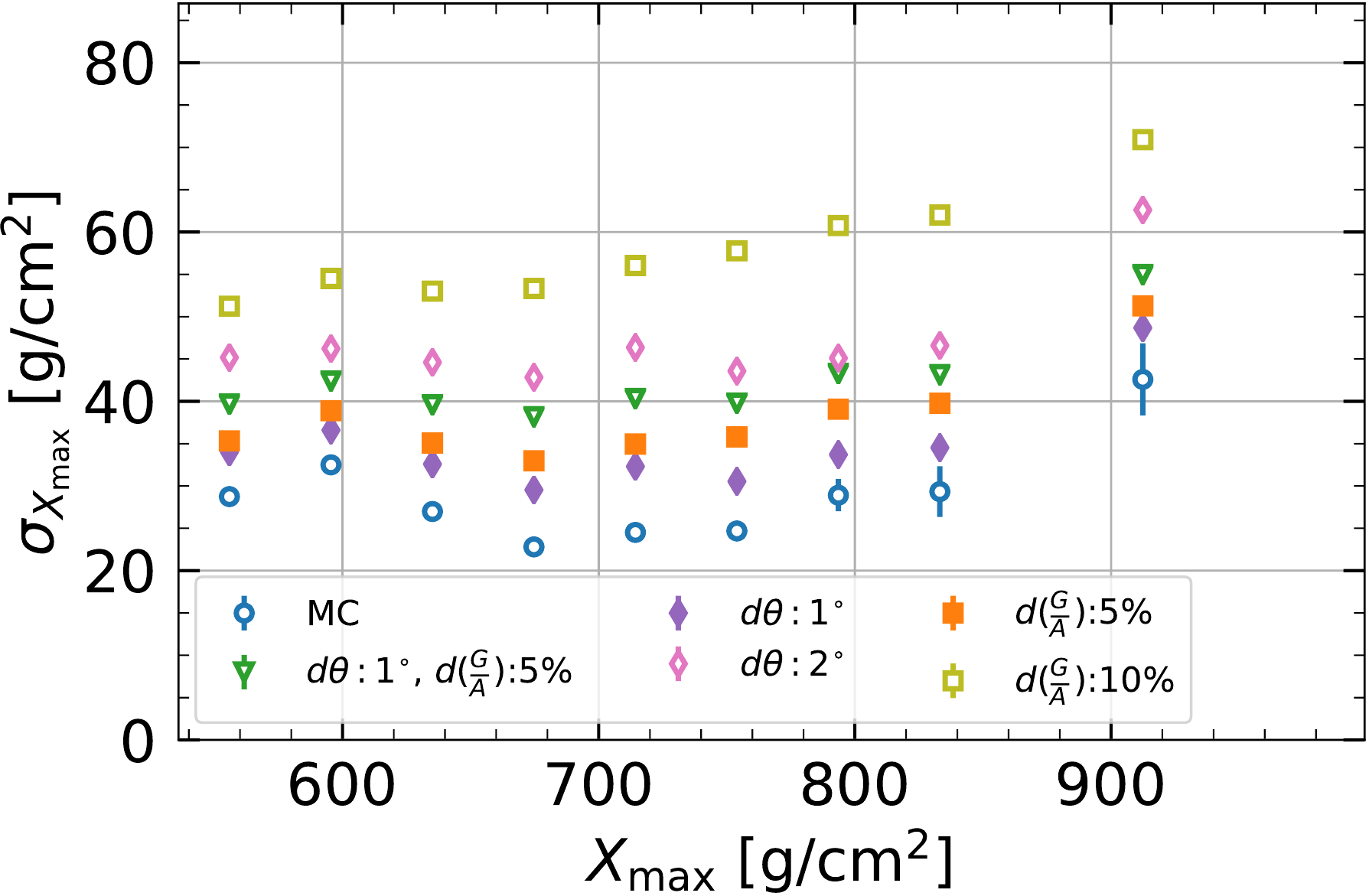}
\caption{Resolution of \cref{eq:dX_pars}, $\sigma_{\Xmax}$, i.e., the spread of the \dX values. Blue circles represent the spread without any experimental uncertainty. Purple and pink diamonds represent a $1^{\circ}$ and $2^{\circ}$ measurement uncertainty in arrival direction. The orange and yellow squares represent a $5\,\%$ and $10\,\%$ measurement uncertainty of $\frac{G}{A}$. The green triangle shows the combination of a 1$^{\circ}$ and $5\,\%$ uncertainty.}
\label{fig:sigmaPlots}
\end{figure}

Using \cref{eq:dX_pars}, we quantify the resolution with which one can identify \Xmax, by $\sigma_{\Xmax}$, the standard deviation of the simulated \Xmax from the \Xmax reconstructed from the model.
The three plots in \cref{fig:sigmaPlots} show the resolution on \Xmax, $\sigma_{\Xmax}$, as a function of \aBarInv, zenith angle, and \Xmax.
Note that the resolution of \dX is equivalent to that of \Xmax, assuming that the atmosphere is known.
The average resolution of the fit for exact Monte-Carlo values is about 35\,\gcm across the entire data set.
However, there are regions of the phase space where the resolution is better than this, in particular, for high zenith angles where $\sigma_{\Xmax} \simeq 20$\,\gcm.

To estimate the uncertainty that would be achievable by an air-shower array, the true values of \aBarInv and \sect for each event, individually, were smeared 100 times and inserted into \cref{eq:dX_pars}.
The resolution of the arrival direction for the IceTop Surface Enhancement at the relevant energies for radio detection ( $\gtrsim$\,30\,PeV) is expected to be $<$\,1$^\circ$~\cite{IceCube:2021qnf,IceCube:2012nn,agnieszka2019scintsim}, and also other radio arrays have demonstrated sub-degree direction accuracy for air showers~\cite{Corstanje:2014waa,LOPES:2021ipp}.
To be conservative, we show the expected resolution assuming a Gaussian resolution of $1^\circ$ and $2^\circ$ (purple and pink diamonds).
The values of $\sigma_{\Xmax}$ are about 5\,\gcm larger for $1^\circ$, but are 15\,\gcm larger at $2^\circ$.
This indicates that the use of this method will depend on a good arrival direction reconstruction, particularly at high zenith angles.

A corresponding process was performed, assuming a relative uncertainty on the measurement of $\frac{G}{A}$.
The expected measurement accuracy for the Askaryan fraction of an individual air shower is not yet known for the IceTop Surface Enhancement and is not available from previous experiments. It will likely depend on the core resolution, the number of antennas contributing to the measurement and their signal to noise ratio, and the location of those antennas in the shower plane. 
For showers featuring high signal-to-noise ratios, the precision of $\frac{G}{A}$ will be limited by systematic uncertainties, such as gain variations between antennas, environmental effects, and other systematic uncertainties beyond those affecting all antennas in the same way or the overall absolute scale. 
Such individual systematic uncertainties on the radio amplitude are estimate to be of $o(5\,\%)$ \cite{PierreAuger:2015hbf,PierreAuger:2017xgp,Bezyazeekov:2015rpa}. 
At lower SNR, noise will contribute an additional uncertainty.

Hence, we investigated target resolutions of $5\,\%$ and $10\,\%$ in $\frac{G}{A}$, which, if achieved, would make this method comparable with other techniques for determining \Xmax using radio measurements.
The values of $\sigma_{\Xmax}$ after smearing $\frac{G}{A}$ by $5\,\%$ and $10\,\%$ as shown in \cref{fig:sigmaPlots} indicate that \Xmax resolutions of better than $40\,$\gcm are feasible for inclined showers, but only if the Askaryan fraction can be measured with a high accuracy of $\simeq$\,5\,\%.

Finally, to estimate the effect of both measurement uncertainties, we show the resolution of $\sigma_{\Xmax}$ after smearing $\frac{G}{A}$ by 5\% and the zenith angle by $1^\circ$.
This results in a resolution of about $40\,$\gcm for most zenith angles up to about $45^\circ$ and about $35\,$\gcm for more inclined showers.

As a final validation of the model, we applied \cref{eq:dX_pars}, using the values from \cref{tab:dX_pars}, to two other sets of simulated air showers with energies of $10^{16.0}$--$10^{16.1}$\,eV and 10$^{18.0}$--10$^{18.1}$\,eV.
Aside from the energy, the simulations settings were the same as those described previously.
The results for the \Xmax resolution were on average within $\pm 3\,$\gcm and hence compatible with those shown in \cref{fig:sigmaPlots}, indicating that the description of the relationship between \dX and \arel found in this work does not have strong energy dependence.

\section{\label{sec:concl}Conclusion}
The relative Askaryan fraction was investigated using proton and iron showers simulated by \textsc{CORSIKA}/\textsc{CoREAS} for the South Pole location and magnetic field. 
The contributions from the geomagnetic and Askaryan emission were identified by looking at the polarization of the electric field along the \vb axis.
The quantification of the Askaryan fraction in dependence of various shower parameters provides important knowledge for the analysis of measurements with the future IceTop Surface Enhancement. 

The Askaryan fraction was found to be small near the shower axis, which is expected for radially polarized emission, and to be at an approximately constant level around the Cherenkov cone.
Hence, we defined a quantity, \aBar, which was used to describe the average value of the Askaryan fraction for individual showers, based on the location of the Cherenkov ring.
The relationship was studied as a function of various shower parameters and a correlation between \aBar, \dX, and zenith angle was found.
With the goal of finding a description of the \Xmax of individual air showers using measurable quantities, we developed a parameterization of the dependence of the distance to shower maximum on \aBar and \sect.

If the true simulated shower parameters and polarization vectors are known, the model was found to produce an \Xmax resolution of 20--40\,\gcm. 
After including a conservative estimate of the angular resolution of $1^\circ$, we find that an angular resolution of 30--55\,\gcm can be achieved if the measurement of $G/A$ can be determined to within 5\%.
In particular at high zenith angles $\gtrsim$\,$50^\circ$, an \Xmax resolution of about 30\,\gcm can be achieved for showers with clear signals not significantly disturbed by noise, which makes this method competitive with other techniques~\cite{Apel:2014usa,Nelles:2014gma,Tunka-Rex:2015zsa,Corstanje:2021kik} with resolution ranging from 20--40\,\gcm.
Only computing-intensive template matching of the radio amplitudes in individual antennas provides a significantly better resolution of $\sim$\,17--35\,\gcm ~\cite{Buitink:2014eqa,Bezyazeekov:2018yjw}.

We therefore conclude, that it is worth investigating whether using the polarization to derive the Askaryan fraction can be included as additional observables in such template-fitting techniques. 
Several of these other techniques, which include explicit parametrizations of the lateral distribution of signal, the wave front shape, and the spectral slope, can be used in conjunction with the method described here to improve the resolution of \Xmax further. 
Machine-learning techniques such as neural networks may provide a tool to further investigate the \Xmax resolution achievable by such multi-variate approaches.


\appendix
\section{\label{appen:phaseShift} Effect of the phase offset between the Askaryan and the geomagnetic emission}

\begin{figure}[tb]
    \centering
    \includegraphics[width=8.6cm]{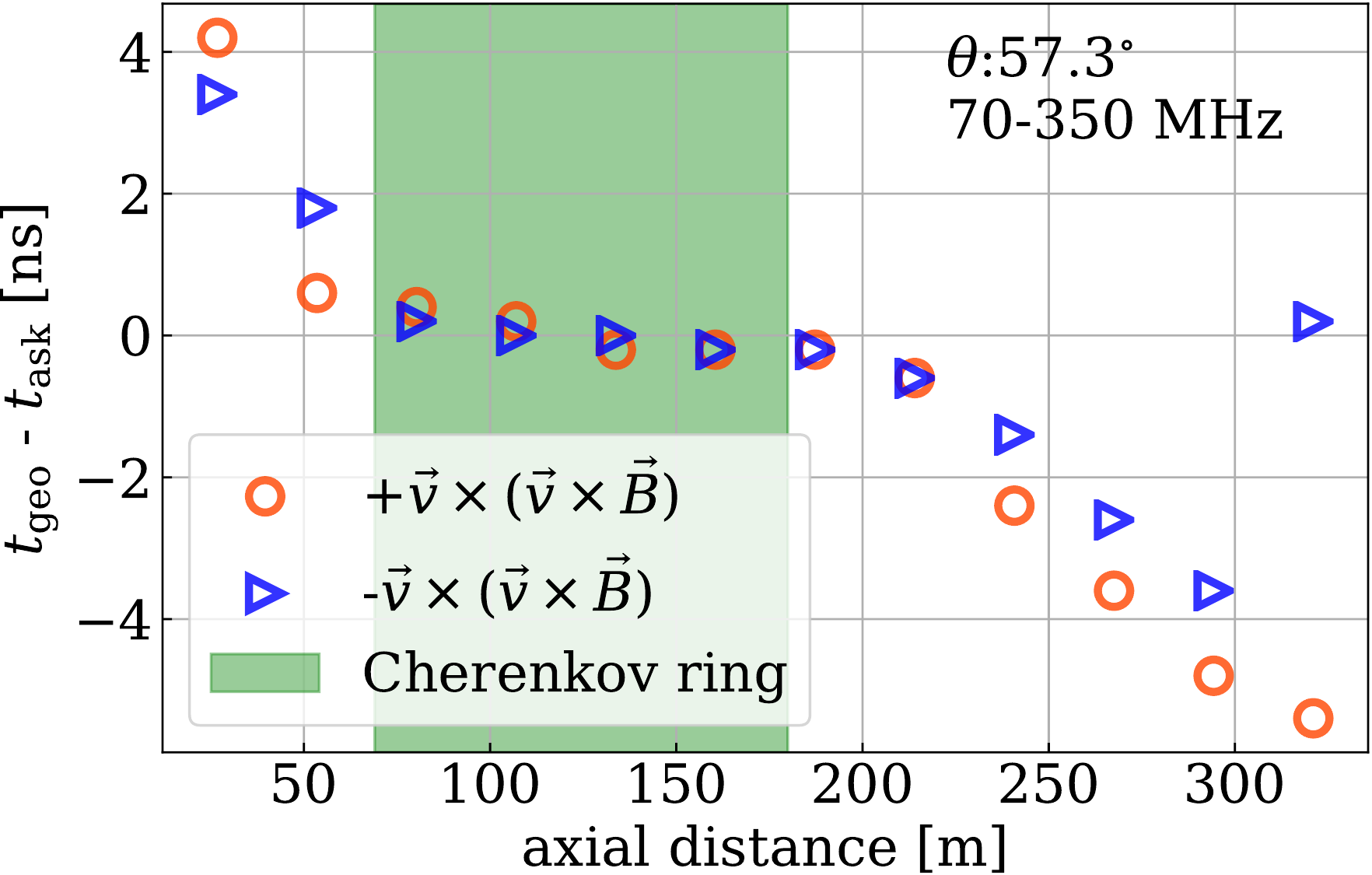}
    \caption{The difference between the Hilbert peak time of the geomagnetic and the Askaryan emission. The location of window, based on the Cherenkov ring (green), is highlighted in green.} 
    \label{fig:peakHtiming}
\end{figure}

The Geomagnetic and Askaryan emissions may not necessarily be in phase with each other due to the difference in how they are produced within an air shower~\cite{Scholten:2016gmj,Glaser:2016qso}. The phase difference  also depends on the  location in the shower footprint where the observation is being made. In  figure \ref{fig:peakHtiming}, the time difference in peak of the geomagnetic emission and Askaryan emission is plotted for various distances from the shower axis. 
The Askaryan emission leads close to the shower axis and lags at radii outside of the Cherenkov ring (shown in green), with differences on the order of a few nanoseconds.
Within the Cherenkov ring, the two peak simultaneously.
Since the method described in this paper only uses the polarization for locations within the Cherenkov ring, the approximation that the peak of the 3D Hilbert envelope describes the peak of the two emission mechanisms is valid.

\begin{figure}[tb]
    \centering
    \includegraphics[width=8.6cm]{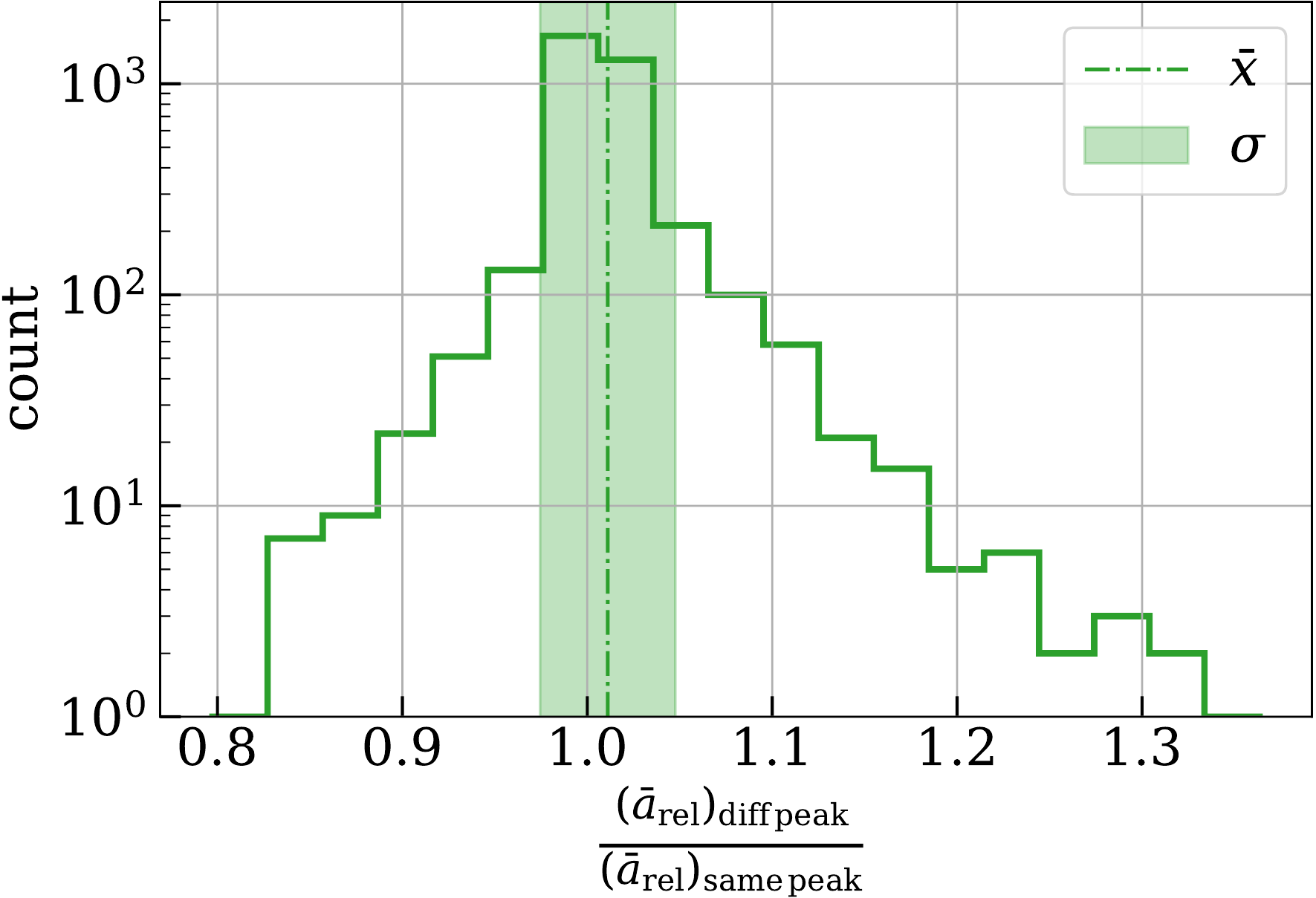}
    \caption{Histogram of the ratio of \aBar calculated without assuming concurrent peak with the value calculated assuming that geomagnetic and Askaryan emission peaks at the same time. The vertical dotted line and the shade represents mean and standard deviation of the ratio.} 
    \label{fig:afWDistance}
\end{figure}

When we compared the plateau value of \aBar at a distance window based on the location of the Cherenkov ring calculated using the emission components at the common peak of the signal with the one calculated using the components at the individual peak of the geomagnetic and the Askaryan emission, their ratio is very close to 1 with mean $\sim$\,1.01 as seen in \cref{fig:afWDistance}.
We thus conclude that we can neglect the phase difference between geomagnetic and Askaryan emission in this study.
\section{\label{appen:alternateMethod} Alternative method of calculating the relative Askaryan fraction}
For the sampling locations along the \vb axis, as can be seen in the figure \ref{fig:antennaMap}, the geomagnetic emission and the Askaryan emission are either aligned along the same direction superimposing constructively or aligned along opposite directions superimposing destructively.
This gives rise to an asymmetry of the lateral distribution of the radio amplitude along the \vb axis. 
To separate the individual contribution of the geomagnetic and the Askaryan emission, we can use the strength of the asymmetry between sampling locations equidistant from the shower axis along the positive and negative \vb axes.
\[E_{+\vb} = G + A\]
\[E_{-\vb} = G - A\]
\begin{equation}
\label{eq:AFvB}
 a_{rel} = \sin \, \alpha \left\lvert\frac{E_{+\vb}-E_{-\vb}}{E_{+\vb}+E_{-\vb}}\right\rvert
\end{equation}
Here, $E_{+\vb}$ is the maximum of the Hilbert envelope of the electric field along the \vb axis at a,certain distance from the shower axis along the positive \vb axis, and $E_{-\vb}$ is the corresponding value along the negative \vb axis. 

\begin{figure}[htbp]
    \subfloat[\centering]{\includegraphics[width=8.6cm]{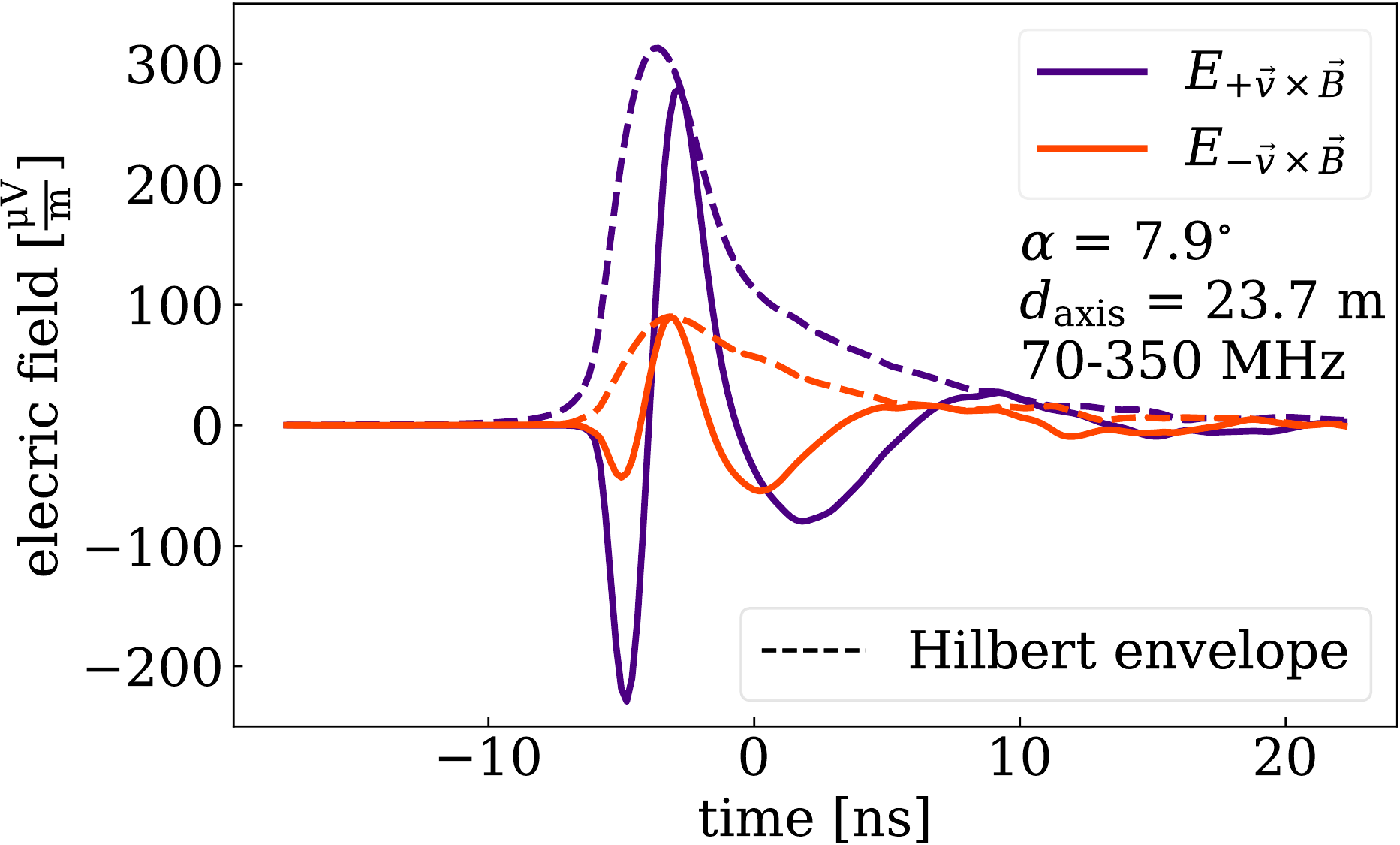}\label{fig:sameSwing}}
    \qquad
    \subfloat[\centering]{\includegraphics[width=8.6cm]{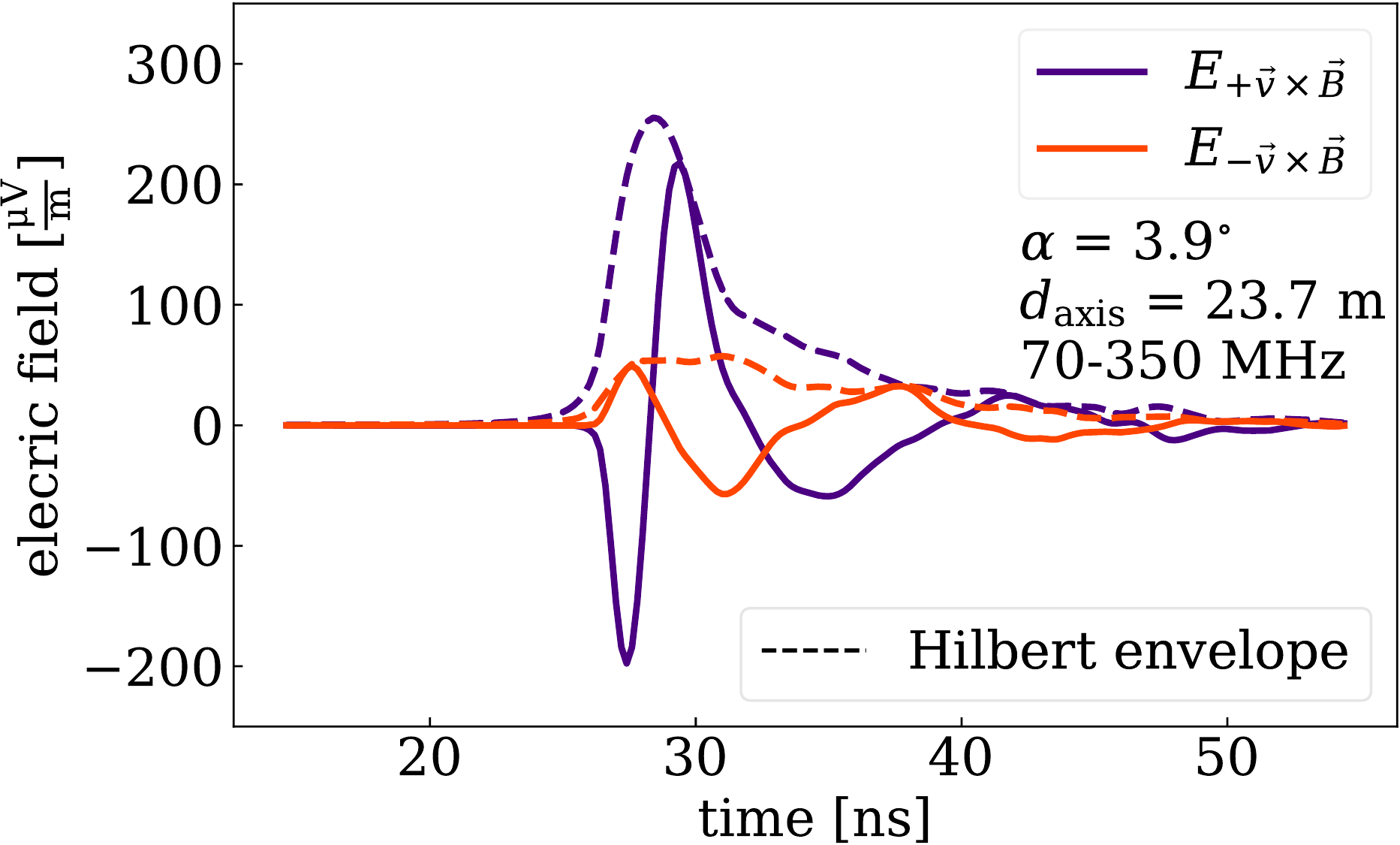}\label{fig:oppoSwing}}
\caption{Trace of the simulated electric field at two sampling locations $\sim$ 24\,m from the shower axis along the positive and negative \vb axes for two simulated air showers: \protect\subref{fig:sameSwing} with geomagnetic angle $\alpha = 7.9^{\circ}$ (geomagnetic emission dominates), and \protect\subref{fig:oppoSwing} with $\alpha = 3.9^{\circ}$ (Askaryan emission dominates).}
\label{fig:swing}
\end{figure}

The relative phase between the electric fields at the locations along the positive and negative \vb axes depends on whether the Askaryan or geomagnetic emission dominates (figure \ref{fig:swing}) and, thus, changes at a geomagnetic angle of about $5^\circ$. 
%
To account for this, the peak Hilbert amplitude $E_{-\vb}$ is assigned a negative sign in the equation \ref{eq:AFvB} to calculate \arel for cases when the Askaryan emission dominates.

\begin{figure}[tb]
    \includegraphics[width=8cm]{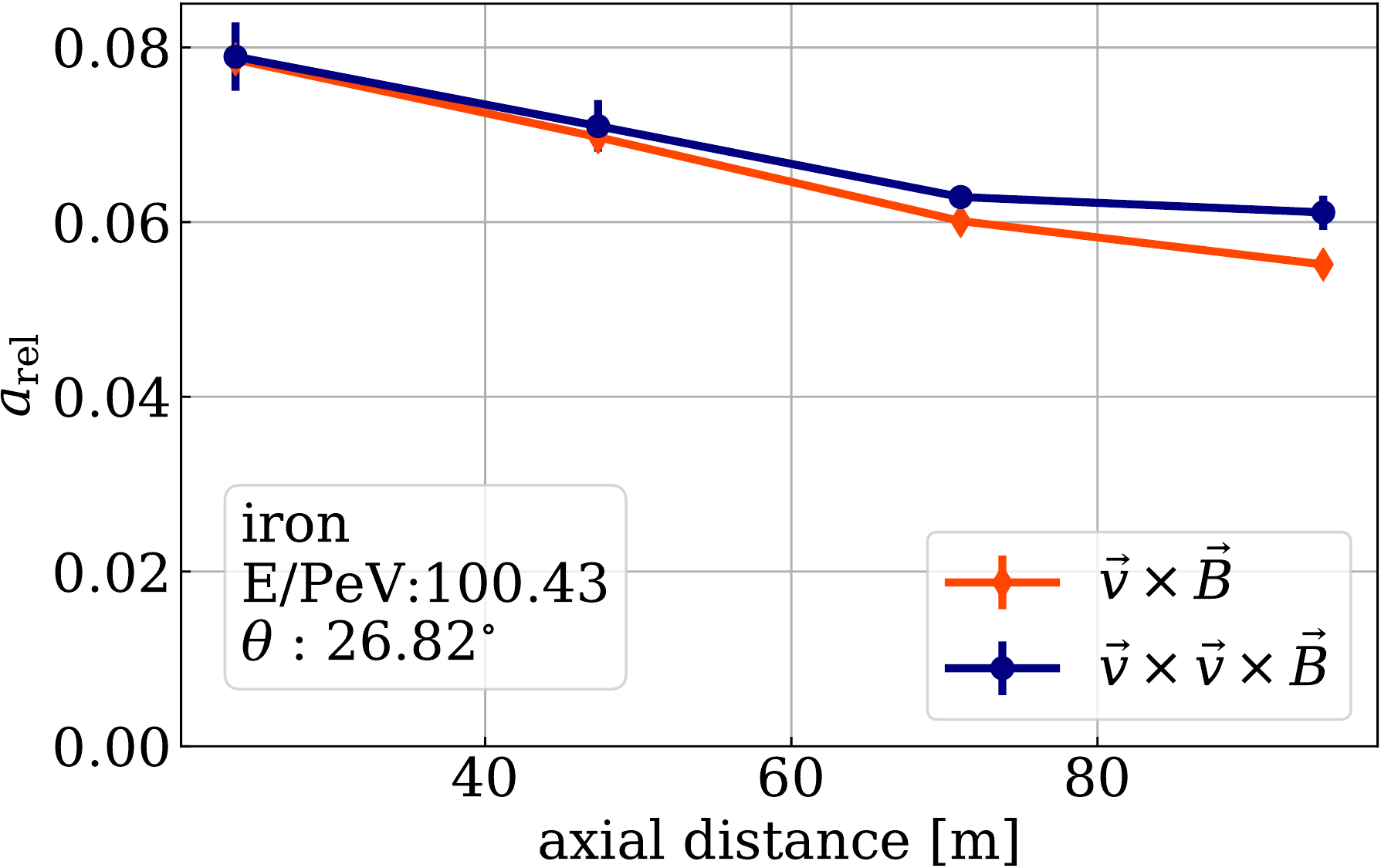}
    \qquad
    \includegraphics[width=8cm]{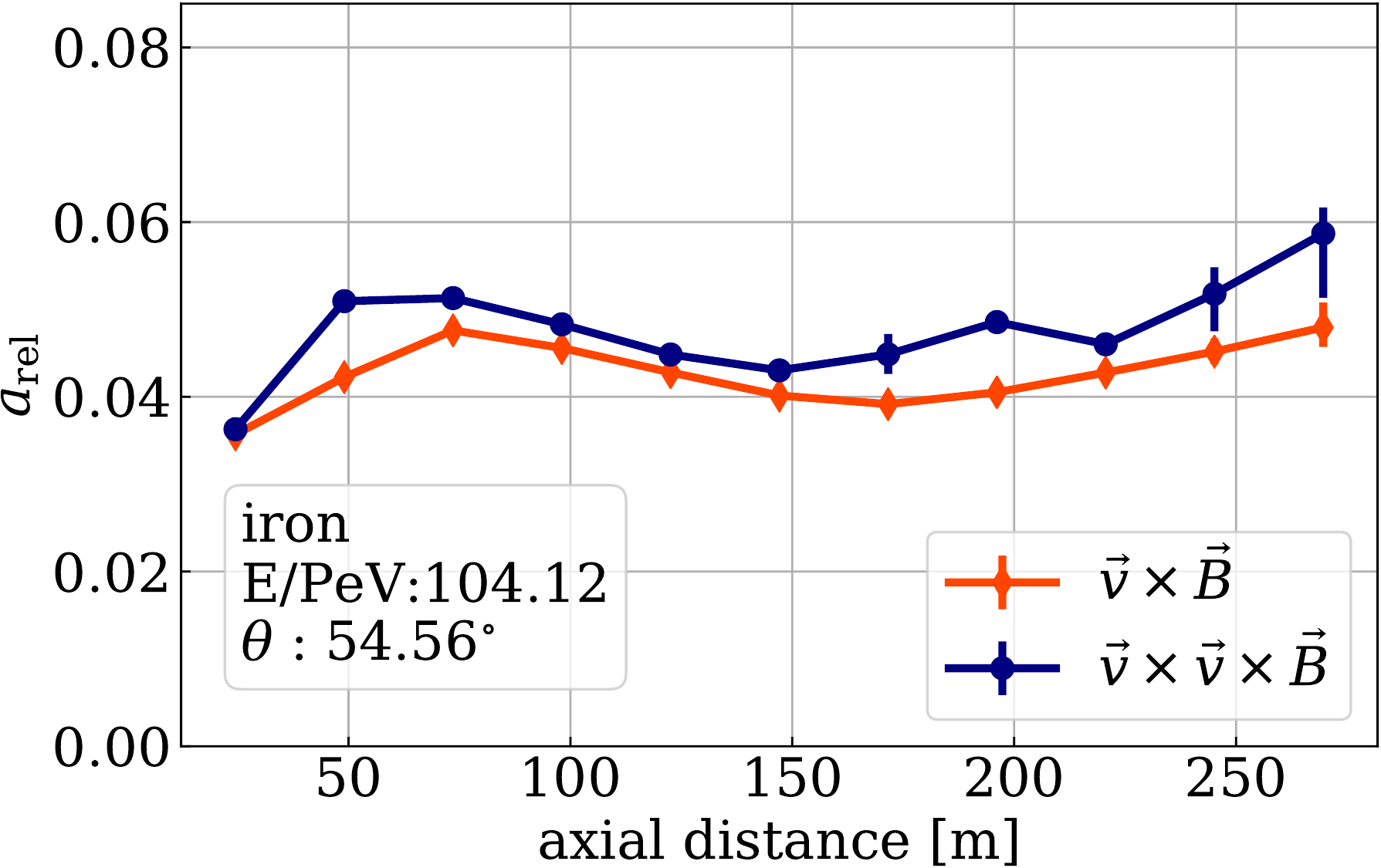}
\caption{Relative Askaryan fraction against the distance from the shower axis using the polarization method along the \vvb axis (blue) and using the asymmetry method along the \vb axis (orange) for two showers with different zenith angles.}
\label{fig:caseTwoAxis}
\end{figure}

Generally, the method of asymmetry and the polarization method for the Askaryan fraction yield consistent results as shown in figure \ref{fig:caseTwoAxis}. 
However, the asymmetry method comes with an intrinsic difficulty. 
Especially for showers with a geomagnetic angle close to $5^\circ$, i.e., at the transition from geomagnetic dominance to the Askaryan dominance, it is difficult to separate the cases shown in figure \ref{fig:swing}. 
Hence, we have decided to use the polarization method for the calculation of the Askaryan fraction, which is applicable for all geomagnetic angles in the same way.

\section{Correlation Matrix}
\label{app:correlation_matrix}
The values of the symmetric correlation matrix corresponding to the fit parameters in \cref{tab:dX_pars} is given below.
{\footnotesize
$$
\left(
    \begin{array}{cccccccc}
        1 & -0.630 & -0.512 & \phantom{-}0.339 & \phantom{-}0.158 & \phantom{-}0.223 & -0.011 & - 0.249\\
         & 1 & -0.195 & -0.494 & \phantom{-}0.049 & -0.113 & \phantom{-}0.189 & \phantom{-}0.063\\
         &  & 1 & -0.212 & -0.449 & \phantom{-}0.022 & -0.109 & \phantom{-}0.273\\
         &  &  & 1 & -0.240 & -0.369 & \phantom{-}0.050 & \phantom{-}0.039\\
         &  &  &  & 1 & -0.156 & -0.400 & \phantom{-}0.099\\
         &  & \cdots &  &  & 1 & -0.198 & -0.492\\
         &  &  &  &  &  & 1 & -0.580\\
         &  &  &  &  &  &  & 1\\
    \end{array}
\right)
$$
}
\bibliography{main}
\end{document}